\documentclass[preprint,superscriptaddress,groupedaddress,unsortedaddress,runinaddress,frontmatterverbose, nofootinbib,nobibnotes,amsmath,amssymb,aps,pra,floatfix,showkeys,
]{revtex4-2}
\pdfoutput=1
\usepackage{graphicx}
\usepackage{bm}
\usepackage{hyperref}
\usepackage{xcolor}
\hypersetup{
    colorlinks=true,
    linkcolor=blue,
    filecolor=blue,      
    urlcolor=blue,
    citecolor=blue,
    hypertexnames=true,
}

\begin{document}
\preprint{}
\title{\textcolor{blue}{Optical Micromanipulation of Soft Materials: Applications in Devices and Technologies}}

\author{Sanatan Halder}
\affiliation{Department of Physics, Indian Institute of Technology Kanpur, Kanpur–208016, India}
\author{Debojit Chanda}
\affiliation{Department of Physics, Indian Institute of Technology Kanpur, Kanpur–208016, India}
\author{Dibyendu Mondal}
\affiliation{Department of Physics, Indian Institute of Technology Kanpur, Kanpur–208016, India}
\author{Sandip Kundu}
\affiliation{Department of Physics, Indian Institute of Technology Kanpur, Kanpur–208016, India}
\author{Manas Khan}
\email{mkhan@iitk.ac.in}
\affiliation{Department of Physics, Indian Institute of Technology Kanpur, Kanpur–208016, India}

\begin{abstract}
 
Since its invention by Arthur Ashkin and colleagues at Bell Labs in the 1970s, optical micromanipulation, also known as optical tweezers or laser tweezers, has evolved remarkably to become one of the most convenient and versatile tools for studying soft materials, including biological systems. Arthur Ashkin received the Nobel Prize in Physics in 2018 for enabling these extraordinary scientific advancements. Essentially, a focused laser beam is used to apply and measure minuscule forces from a few piconewtons to femtonewtons by utilizing light-matter interaction at mesoscopic length scales. Combined with advanced microscopy and position-sensing techniques, optical micromanipulations enable us to investigate diverse aspects of functional soft materials. These include studying mechanical responses through force-elongation measurements, examining the structural properties of complex fluids employing microrheology, analyzing chemical compositions using spectroscopy, and sorting cells through single-cell analysis. Furthermore, it is utilized in various soft-matter-based devices, such as laser scissors and optical motors in microfluidic channels. This chapter presents an overview of optical micromanipulation techniques by describing fundamental theories and explaining the design considerations of conventional single-trap and dual-trap setups as well as recent improvisations. We further discuss their capabilities and applications in probing exotic soft-matter systems and in developing widely utilized devices and technologies based on functional soft materials.

\end{abstract}
\maketitle

\newpage

\tableofcontents


\section{Introduction}

The ability of light to exert force on matter was first recognized by Kepler in the early 17th century, as he described that the tails of comets are deflected in response to the light pressure produced by the sun’s light \cite{Kepler1619}. However, it was in the late 19th century that the radiation pressure that acts along the direction of light propagation was understood, describing light propagation through Maxwell’s theory of electromagnetism \cite{Poynting1884}. In initial experiments, Nicholas and Hull \cite{Nichols1901} and Lebedew \cite{Lebedew1901} were able to detect the effect of radiation pressure of light on macroscopic objects using arc lamps as the light source and a torsion balance for detection. Intense and coherent light from lasers \cite{Siegman1986} makes it possible to generate sufficiently large forces that are relevant to many microscopic phenomena in physics and biology. Moreover, advancements in various microscopy and imaging techniques, such as confocal imaging and two-photon excitation, have made it possible to visualize and track the dynamics of smaller objects, even up to a few nanometers. In a seminal contribution, Arthur Ashkin and co-workers first demonstrated that a laser beam can be used to hold, that is, to trap and manipulate particles \cite{Ashkin1970,Ashkin1986}. In a series of subsequent experiments, they demonstrated the optical micromanipulation of atoms \cite{Ashkin1970a,Chu1985,Chu1986}, and biological matter \cite{Ashkin1987,Ashkin1987a}. Within a few years, Block \textit{et. al.} used an optical trap to apply forces and thus manipulate E. coli ﬂagella \cite{Block1989} and single kinesin motors \cite{Block1990}. It was then accepted that optical tweezers provide the most ﬂexible tool that uses light to noninvasively trap and manipulate microscopic objects for ultra-fine positioning, measurement, and control. Over time, it has become the most popular and convenient tool for studying soft materials and biological systems, as it provides more powerful yet more convenient ways to probe and manipulate these systems. Arthur Ashkin was later awarded the \textit{2018 Nobel Prize in Physics} (half of the prize) for his invention of optical tweezers \cite{AshkinNobel}.

\begin{figure*}[ht]
	\centering
	\includegraphics[width=1\linewidth]{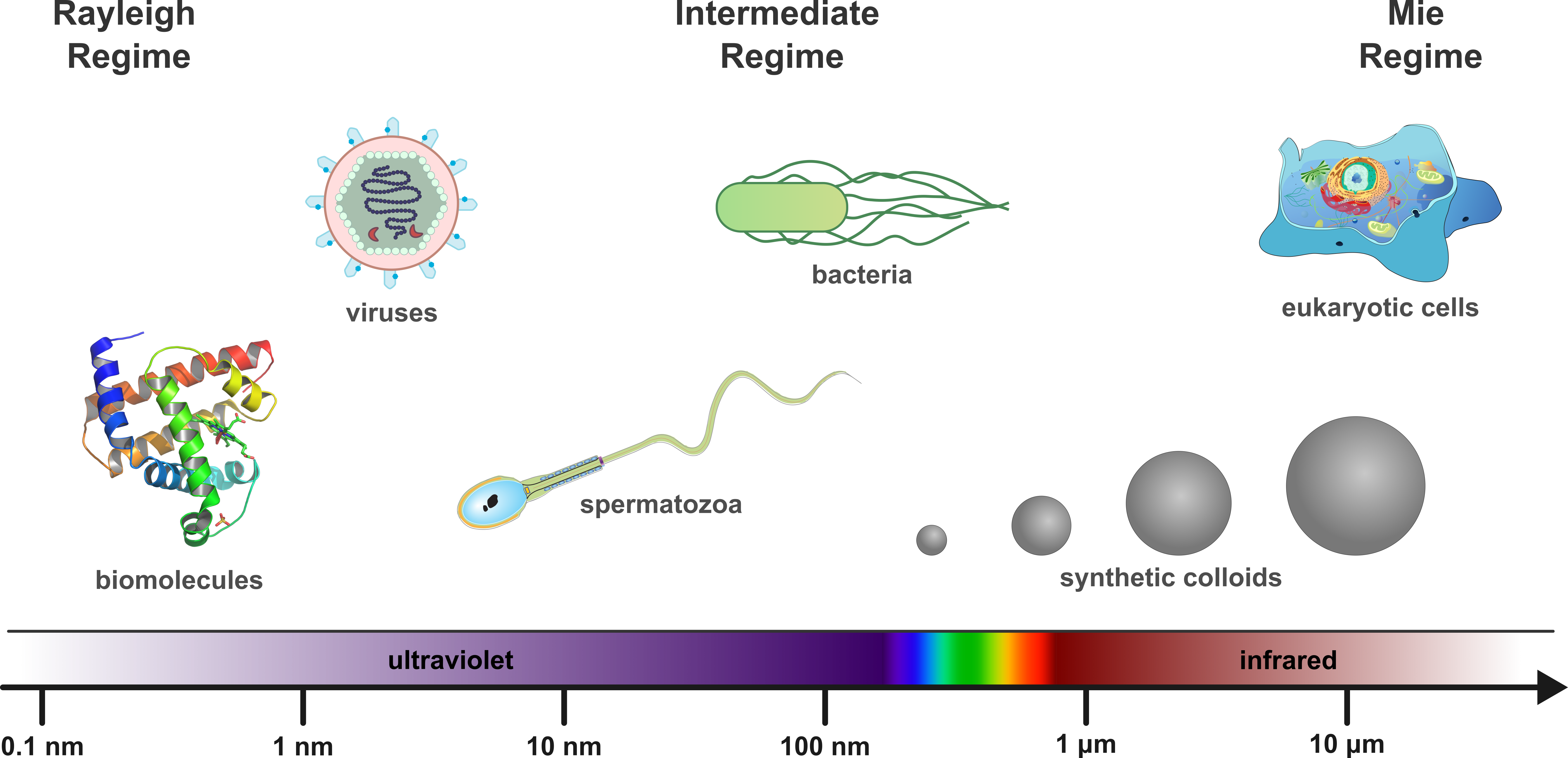}
	\caption{Typical examples of systems trapped and manipulated using optical tweezers in different regimes based on the size of the object with respect to the wavelength of the light beam used. If the object size is smaller or greater than the wavelength, it falls within the Rayleigh or Mie regime. In the intermediate regime, the size of the object is comparable to the wavelength of the laser. Trapping lasers are mostly used in the visible or near-infrared spectral regions.}
	\label{fig:LS}
\end{figure*}

In their simplest configuration, optical tweezers are set up with a single laser beam tightly focused down to the diffraction limit, thereby creating the steep light intensity gradients required to hold dielectric particles in three dimensions at the point of focus. Over the years, this conventional form has evolved with many improvisations and modifications to provide sophisticated and versatile tools for holding, manipulating, probing, and sorting particles, starting from single molecules to entire mammalian cells and colloidal particles (Fig. \ref{fig:LS}) in diverse environments, such as liquids, complex fluids, and even air. Currently, there are multiple available optical micromanipulation techniques, including optical tweezers, optical scissors, and optical spanners. The domain of their applicability has widened substantially. Employing these tools, forces of up to $\sim$200 picoNewton can be applied with sub-picoNewton resolution. The emerging applications being quite diverse and extensive, optical tweezers have been a focal point of interdisciplinary science, with revolutionary applications in soft materials, including biological systems \cite{Block1990,Ashkin1990,Bustamante1994,Finer1994,Bustamante2003,Neuman2008}. It has also proven useful in various other areas of physics \cite{Grier2003,Dholakia2008,Jonas2008,Khan2010,Khan2011,Khan2011a, Padgett2011,Padgett2011a,Dholakia2011,Juan2011,Polimeno2018}, nanotechnology \cite{Khan2006,Khan2006a,Khan2007,Marago2013}, and chemistry \cite{Yi2006}. Many interesting review articles and books have been written on optical trapping and manipulation, their design, and their applications  \cite{Simpson1996,Berns1997,Grier1997,Sheetz1998a,Greulich1999,Grier2003,Dholakia2008,Jonas2008,Zhang2008,Padgett2011,Padgett2011a,Dholakia2011,Juan2011, Khan2014, Gennerich2017, Polimeno2018,Pesce2020,Bustamante2021,Yang2021, Volpe2023}.

This chapter cannot cover the full extent or diversity of the applications of optical micromanipulation techniques in soft matter-based systems and devices, but hopes to provide an introduction to optical micromanipulation, including the fundamental phenomena of its operations, some of the latest advancements, its principal capabilities, and finally a few important applications. It is organized as follows: in the next section we discuss optical micromanipulation techniques including the theory of optical trapping at two different regimes, considerations and design of the conventional and various contemporary  realizations of optical tweezers. Next, we explain the most utilized capabilities of optical tweezers. Finally, some of the important applications of optical micromanipulation in soft-material-based systems, devices, and technologies are discussed.

\section{Optical Micromanipulation Techniques}
Optical micromanipulation techniques enable us to hold and precisely control the movement of micron and sub-micron sized particles using a laser beam. The basic principle that this relatively contemporary tool relies on is the interaction between the electromagnetic field of the light beam and a dielectric particle. A focused laser beam, with higher energy density is required to enhance the interaction, and thus enables successful optical micromanipulation. For simplicity, let us first consider the change in momentum of the photons when a light beam passes through an interface between two materials with different refractive indices. The photons change their speed and direction of propagation, and hence the momentum, upon crossing the interface. This difference in the momentum generates a finite force, which is of the order of pico-Newton and exerted on to the interface. If the second medium is a micron sized particle suspended in the first medium, the force exerted by this means can become significant to manipulate the particle.

In 1969, Arthur Ashkin first showed that it is possible to accelerate a micron sized dielectric particle, suspended in liquid, using a laser beam \cite{Ashkin1997}. A year later, in 1970, Ashkin reported that two counter propagating laser beams can apply precisely (un)balancing forces to hold and manipulate such particles, and water droplets in air. He also predicted the extension of optical manipulation to atoms and molecules \cite{Ashkin1970}. Finally, in 1986, Ashkin and colleagues could successfully trap dielectric particles with size ranging from $\sim$ 25 nm to 10 $\mu$m with a tightly focused single laser beam \cite{Ashkin1986}. This technique became popular as {\it{Optical Tweezers}} and has been proven to be very convenient and useful in various fields of research and applications, particularly with soft materials. Arthur Ashkin was awarded the Nobel prize in Physics in 2018 for inventing the {\it{Optical Tweezers}} and their application to biological systems.

\subsection{Theory of Optical Micromanipulation}

When a dielectric particle enters the electromagnetic field of a laser beam and interacts with it, both electric field strength and polarization play important roles. However, for simplicity, we consider a linearly polarized laser beam and nonbirefringent particles. The force exerted by the laser field on the dielectric particle can be decomposed into two principal components of different origins: gradient force and scattering force. The scattering of the laser beam from the particle gives rise to a scattering force, which acts along the direction of propagation of the beam if it has a symmetric intensity profile. Thus, the scattering force pushes the particle along, inducing a destabilizing effect. In contrast, the gradient force is conservative in nature and stabilizes the particle by pushing it towards the point of focus. As the name suggests, the gradient force field is proportional to the spatial gradient of intensity. Therefore, a beam with a Gaussian lateral intensity profile ($TEM_{00}$ mode) creates a restoring force field centered at the point of maximum intensity, that is, the point of focus. The origin of the gradient force can be explained in two different ways depending on the size of the particle. For particles that are larger than the wavelength of the trapping laser, simple ray optics principles, such as the refraction of the light beam through the particle, can explain the gradient force in the Mie regime. In this regime, the transparency of the particles at the trapping wavelength plays an important role. Particles with higher transmittance experience a dominant gradient force, whereas the scattering force is stronger when the reflectance is greater. In the Rayleigh regime, where the particle size is smaller than the wavelength of the trapping laser, we must consider the particle as a Rayleigh scatterer and use Maxwell’s laws of electromagnetism to understand the origin of the gradient force.

\subsubsection{Mie Regime}

\begin{figure*}[ht]
	\centering
	\includegraphics[width=0.75\linewidth]{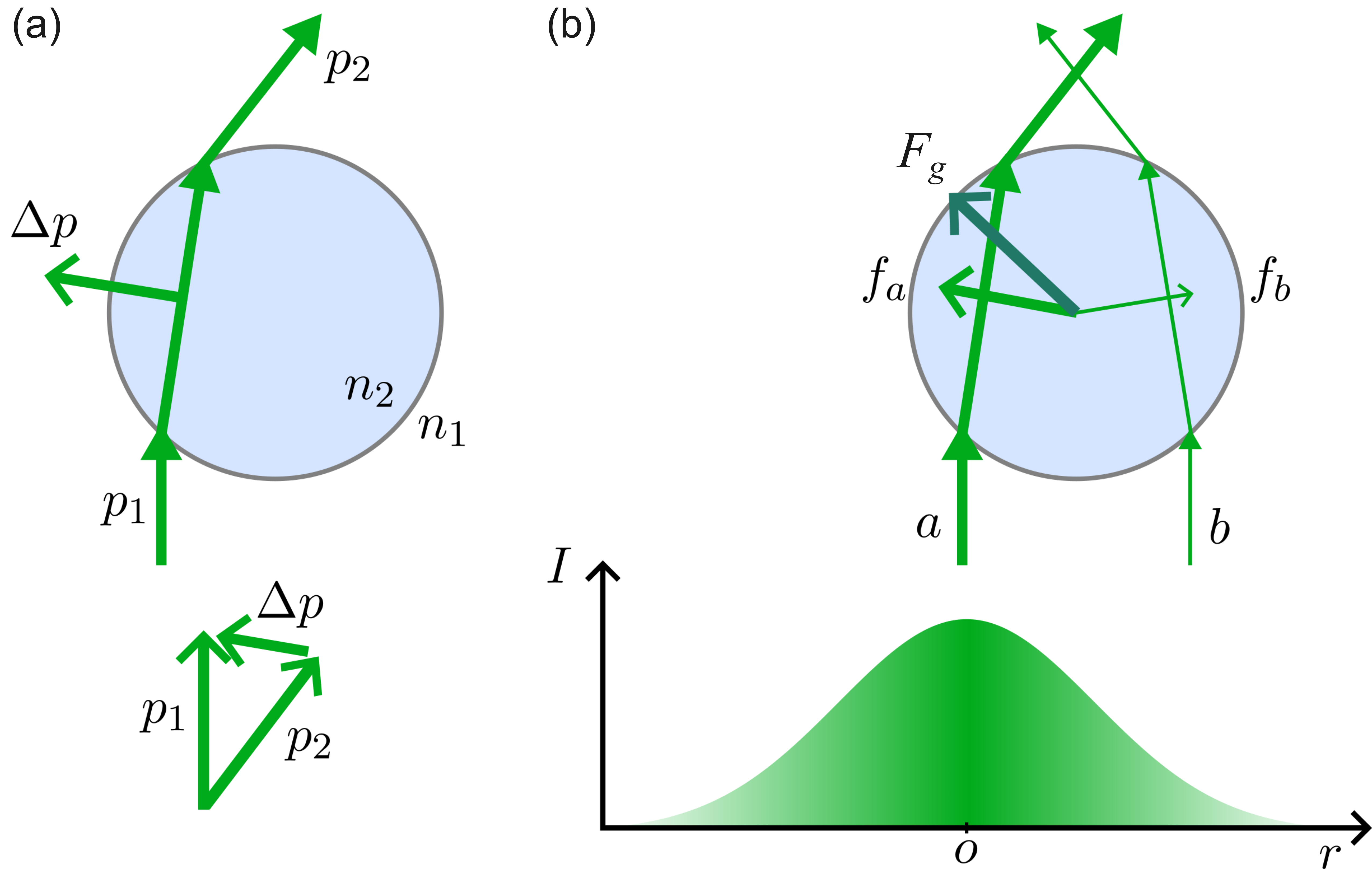}
	\caption{Origin of the gradient force in Mie regime. (a) A ray with momentum $p_1$ gets transmitted through a micro-particle and emerges with momentum $p_2$. The difference in momentum, $\Delta p$, causes a motion of the particle to conserve the linear momentum. (b) When the rays (two typical ones are marked as $a$ and $b$) transmit through the particle, the change in their momentum gives rise to  forces (e.g., $f_a$ and $f_b$ respectively). Depending on the intensity field of the laser at the position of the particle, different rays that pass through the particle have varied momentum. A ray from the central region of the laser profile (e.g., $a$) will have higher momentum and consequently applies stronger force ($f_a$) to the particle as compared to a ray from the periphery (e.g., $b$ and corresponding force $f_b$). Thus, the net force always acts along the direction of the highest intensity.}
	\label{fig:GFM}
\end{figure*}

In the Mie regime, where the particle diameter is larger than the laser wavelength, diffractive effects can be neglected, and the geometrical optics approach can be safely employed  \cite{Ashkin1992,Visscher1992a,Bustamante2021,Khan2014,Ren1994,Sheetz1998a}. The laser beam is decomposed into rays with appropriate intensities and directions that propagate along straight lines in media with a uniform refractive index. The rays have the characteristics of a plane wave that changes direction upon reflection and refraction, following Fresnel’s equations. When a transparent microparticle is within the spatial intensity gradient of the laser, the refraction of rays with varied intensity through  the particle causes a change in the total momentum of the light beam.  Fig. \ref{fig:GFM} pictorially shows the origin of the gradient force in Mie regime. When an incident ray with momentum $\vec{p_{1}}$ is refracted through the particle and emerges with momentum $\vec{p_{2}}$, the particle experiences a change in momentum of $\Delta\vec{p}$, where $\vec{p_{1}} = \vec{p_{2}} + \Delta\vec{p}$, according to the conservation of linear momentum. Based on the position of the particle in the laser intensity field, rays with varied intensities, and thus different momenta, pass through the particle, exerting forces accordingly. Therefore, the sum of the forces acting on the particle depends on its position. However, the net force is always directed towards the point of highest intensity and pushes the particle towards the point of focus, irrespective of its initial position. This net force, which is proportional to the spatial gradient of light intensity, is called the gradient force.   

\begin{figure*}[ht]
	\centering
	\includegraphics[width=0.65\linewidth]{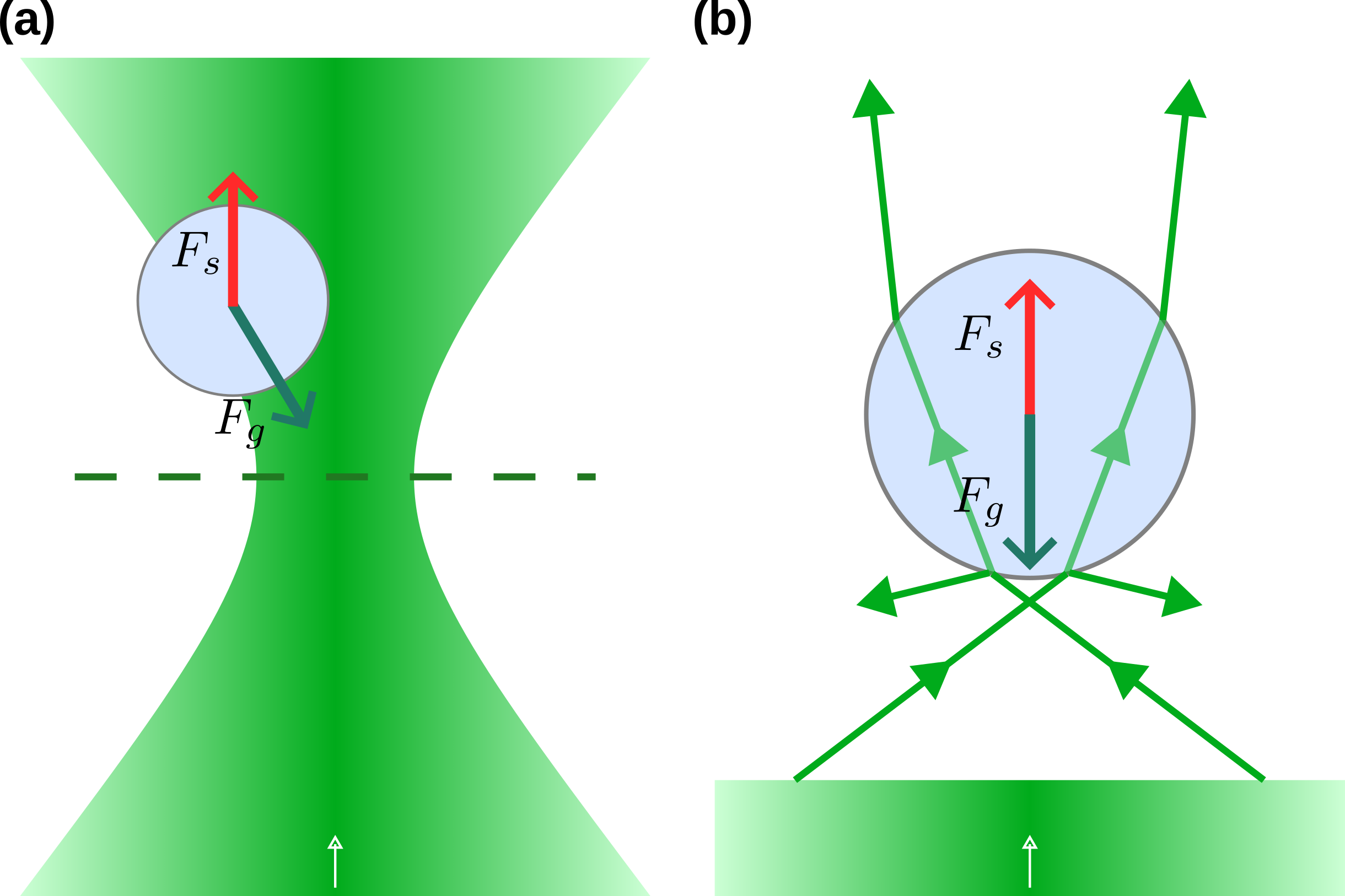}
	\caption{Origin and direction of the scattering force in Mie regime. (a) The directions of the gradient force ($F_g$) and the scattering force ($F_s$) are shown for an off-axis particle. While the gradient force is directed towards the point of maximum intensity, the direction of scattering force remains along the propagation direction of the laser beam. (b) A pair of rays from a focused laser beam is shown to fall on the particle which is situated on the axis, just above the point of focus (where the rays intersect). Since the lateral components balance each other, the net scattering force always acts along the direction of propagation of the laser beam, directly opposing the gradient force.}
	\label{fig:SFM}
\end{figure*}

Some parts of the rays are reflected at the surface of the particle depending on the reflectance of the particle at the wavelength of the trapping laser. The change in momentum upon reflection results in a new force. For the symmetric intensity profile of the laser beam, these forces add up to cancel all the lateral components and generate a force that pushes the particle along the direction of propagation. This phenomenon is known as the scattering force. Fig \ref{fig:SFM} shows the origin and direction of the scattering force. While the gradient force acts along the direction of maximum intensity, and hence the direction changes depending on the position of the particle in the laser intensity field, the scattering force always applies along the propagation direction, irrespective of the position of the particle. The particle in Fig \ref{fig:SFM}(a) is positioned at an off-axis position, whereas in Fig \ref{fig:SFM}(b) it is positioned on axis. In both cases, the scattering force acts vertically downwards, which is the propagation direction, although the gradient force changes its direction.

\subsubsection{Rayleigh Regime}
  
When the particle size is smaller than the wavelength of the trapping laser, the ray optics approach cannot be used as diffraction plays a significant role. Since the perturbation of the incident wavefront is minimal, the particle is considered as a Rayleigh scatterer. The trapping force is then derived using electromagnetic laws \cite{Visscher1992,Harada1996,Sheetz1998a,Khan2014}. When a dielectric particle experiences the electric field associated with the light beam, it becomes polarized because of its finite polarizability. For a particle with positive polarizability, relative refractive index being greater than one, the energy is minimized as it moves to the highest field intensity. This gives rise to the gradient force in the Rayleigh regime and draws the particle toward the point of focus. The net force the particle experiences in this process is the sum of gradient and scattering force. The Lorentz force acting on the induced dipole in the electromagnetic field gives rise to the gradient force. A dielectric particle with polarizability $\alpha$ collects an induced polarization $\vec{p}(\vec{r},t) = \alpha \vec{E}(\vec{r},t)$, where $\vec{E}(\vec{r},t)$ is the electromagnetic field of the laser beam. Therefore, the Lorentz force can be given by,
\begin{eqnarray}
	\vec{F}(\vec{r},t) & = & [\vec{p}(\vec{r},t) \cdot \nabla] \vec{E}(\vec{r},t) + \frac{d\vec{p}(\vec{r},t)}{dt} \times \vec{B}(\vec{r},t)\\
	& = & \alpha \left[ \left( \vec{E}(\vec{r},t) \cdot \nabla \right) \vec{E}(\vec{r},t) + \frac{d\vec{E}(\vec{r},t)}{dt} \times \vec{B}(\vec{r},t) \right] .
\end{eqnarray}
At frequencies much smaller than that of the laser light ($~ 10^{14}$ Hz), the particle experiences a steady force which is time average of $\vec{F}(\vec{r},t)$. This gives the simplified expression for the gradient force, which can be written as, 
\begin{equation}
	\vec{F}_{g}(\vec{r}) = \left\langle \vec{F}(\vec{r}, t) \right\rangle _t = \frac{1}{2} \alpha \nabla \left\langle \vec{E}(\vec{r},t)^2\right\rangle _t = \frac{1}{4} \alpha \nabla \left| E(\vec{r}) \right|^2.
\end{equation}
Relating the intensity of the laser beam $I(\vec{r})$ with $ \left| E(\vec{r}) \right|^2$, using the equation $I(\vec{r}) = (n_1 \epsilon_0 c/2) \left| E(\vec{r}) \right|^2 $, the gradient force simplifies to,
\begin{equation}
	\vec{F}_{g}(\vec{r}) = \frac{1}{2n_1 \epsilon_0 c} \alpha \nabla I (\vec{r}).
\end{equation}
Here $n_1$, $ \epsilon_0$, and $c$ are the refractive index of the medium, permittivity of the free space, and the velocity of light, respectively. Thus, the magnitude of the gradient force is proportional to the gradient of the intensity field, justifying the name.  The direction of the force is always towards the point of highest light intensity, \textit{i.e.}, toward the laser beam axis for a Gaussian beam profile, and toward the point of focus if the laser beam is focused. The potential energy of a dielectric particle in the laser intensity field can then be given by,
\begin{equation}
	U = - \frac{\alpha}{n_1 \epsilon_0 c} \int I(\vec{r}) d^3 r.
\end{equation}
Therefore, to successfully trap a small dielectric particle it needs to have high polarizability.

Another force that arises in this scenario is because of the radiation pressure on the particle, and hence, is called the scattering force. The particle absorbs a part of the incident radiation and scatters it partially. A net force results toward incident photon flux,\textit{ i.e.}, along the propagation of light. This scattering force is given by,
\begin{equation}
	\vec{F}_s = n_1 \frac{\sigma \left\langle \vec{S}\right\rangle_t }{c},
\end{equation}
where $n_1$ is the refractive index of the medium, $c$ is the velocity of light, $\left\langle \vec{S}\right\rangle_t$ is the time-averaged Poynting vector, and $\sigma$ is the scattering cross-section of the particle. The scattering cross-section, $\sigma$ for a spherical particle is well known, and given by,
\begin{equation}
	\sigma = \frac{8}{3} \pi \left( kr \right)^4 r^2 \left( \frac{n^2 - 1}{n^2 + 2}\right)^2 , 
\end{equation}
$k$, $r$, and $n$ are the wave vector of the incident light, radius and the refractive index of the particle, respectively. Thus, the scattering force has a strong particle size dependence, through the scattering cross-section. 

\begin{figure*}[ht]
	\centering
	\includegraphics[width=0.8\linewidth]{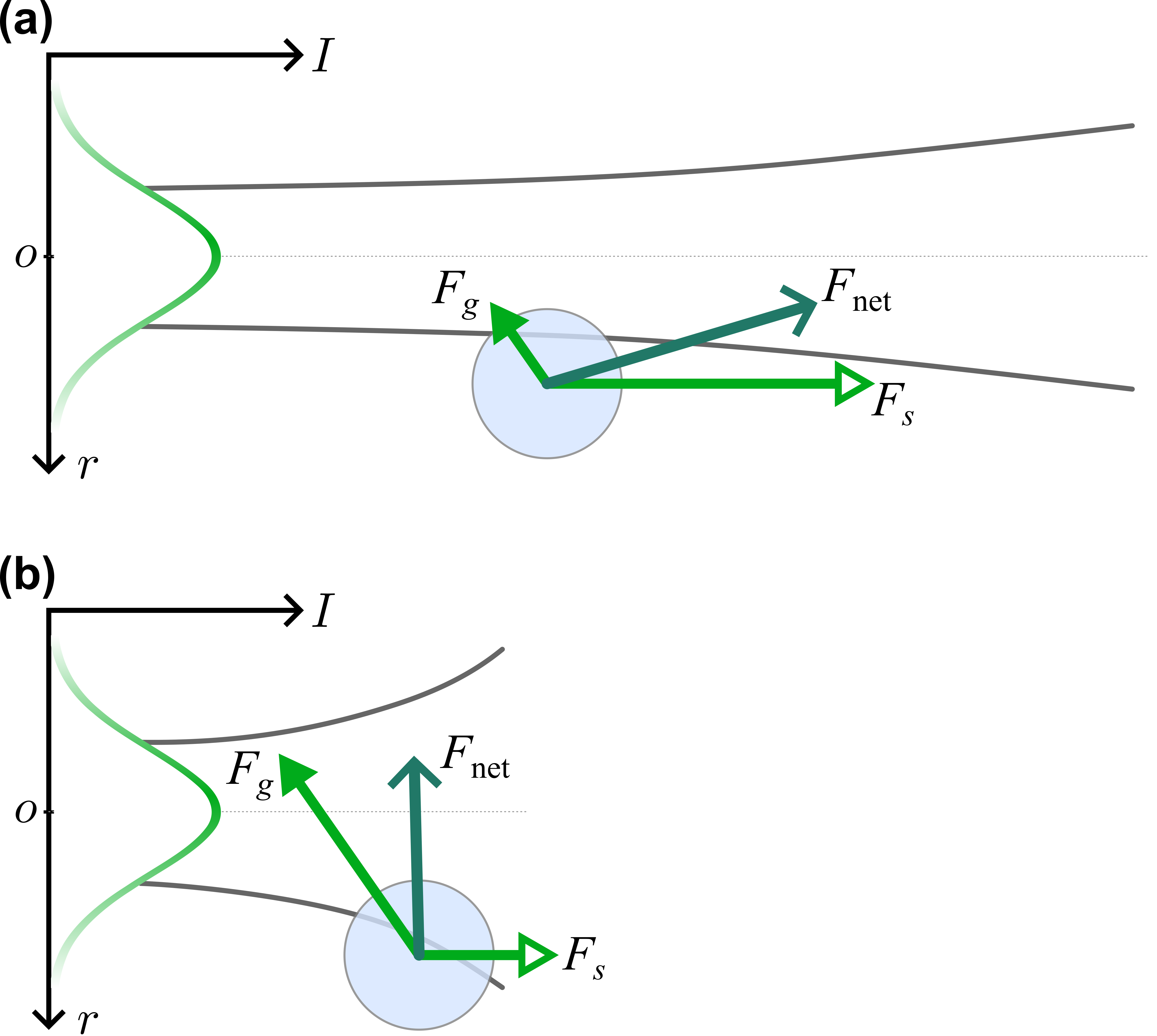}
	\caption{Comparison of relative strength of the gradient force, $F_g$ and the scattering force ($F_s$), for two cases. (a) Low curvature beam, where the scattering force is stronger than the gradient force and the net force ($F_{\textrm{net}}$) pushes the particle away, and hence, the trap becomes unstable. (b) Tightly focused beam, where the component of $F_g$ along the propagation direction of the laser balances the scattering force and thereby stabilizes the particle in the trap.}
	\label{fig:GF}
\end{figure*}

A particle situated away from the beam axis experiences the gradient force as a restoring force towards the beam waist. For a low curvature beam, \textit{i.e.}, a loosely focused laser beam, the component of the gradient force along a direction opposite to that of light propagation becomes too weak to balance a strong scattering force. Hence, the scattering force pushes the particle along the beam. To achieve a stable trap, one requires a counter propagating beam to balance the respective scattering forces \cite{Ashkin1970}. However, in a tightly focused laser beam, the gradient force is directed along the point of focus, and it possesses a component against the Poynting vector, or the scattering force. That component prevents the particle from being pushed away by the scattering force, resulting in a stable confinement of the particle in three dimensions, as is illustrated in Fig. \ref{fig:GF}.

Although the fundamental physics to hold and manipulate a microscopic particle remains the same, it can be realized in various ways, depending on the requirements. Designing a pair of optical tweezers with a Gaussian laser beam is the traditional method, as was shown by Arthur Ashkin in 1986 \cite{Ashkin1986}, and endless customization has been adopted since then. Different beam shapes, such as a Laguerre-Gaussian (LG) beam, which has a dark center resulting in reduced radiation pressure, or a helical beam, have been employed to transfer angular momentum to the particle alongside trapping \cite{He1995,He1995a,Bains1996,Friese1996,Simpson1996,Simpson1997}. New trapping mechanisms such as trapping near a surface with an evanescent wave or near-field enhancement using metallic nanostructures have also been reported \cite{Almaas1995,Friese1998,Friese1998a, Zemanek1998,Fuhr1998,Vasnecov1999,Reicherter1999,Zemanek1999,Clapp1999,Lester1999,Mogensen2000,Schnelle2000,Clapp2001}. In the following section, we discuss the conventional method and some of the more relevant contemporary optical micromanipulation techniques.

\subsection{Conventional Techniques}

\subsubsection{Single Optical Trap}

\begin{figure*}[ht] 
	\centering
	\includegraphics[width=1\linewidth]{"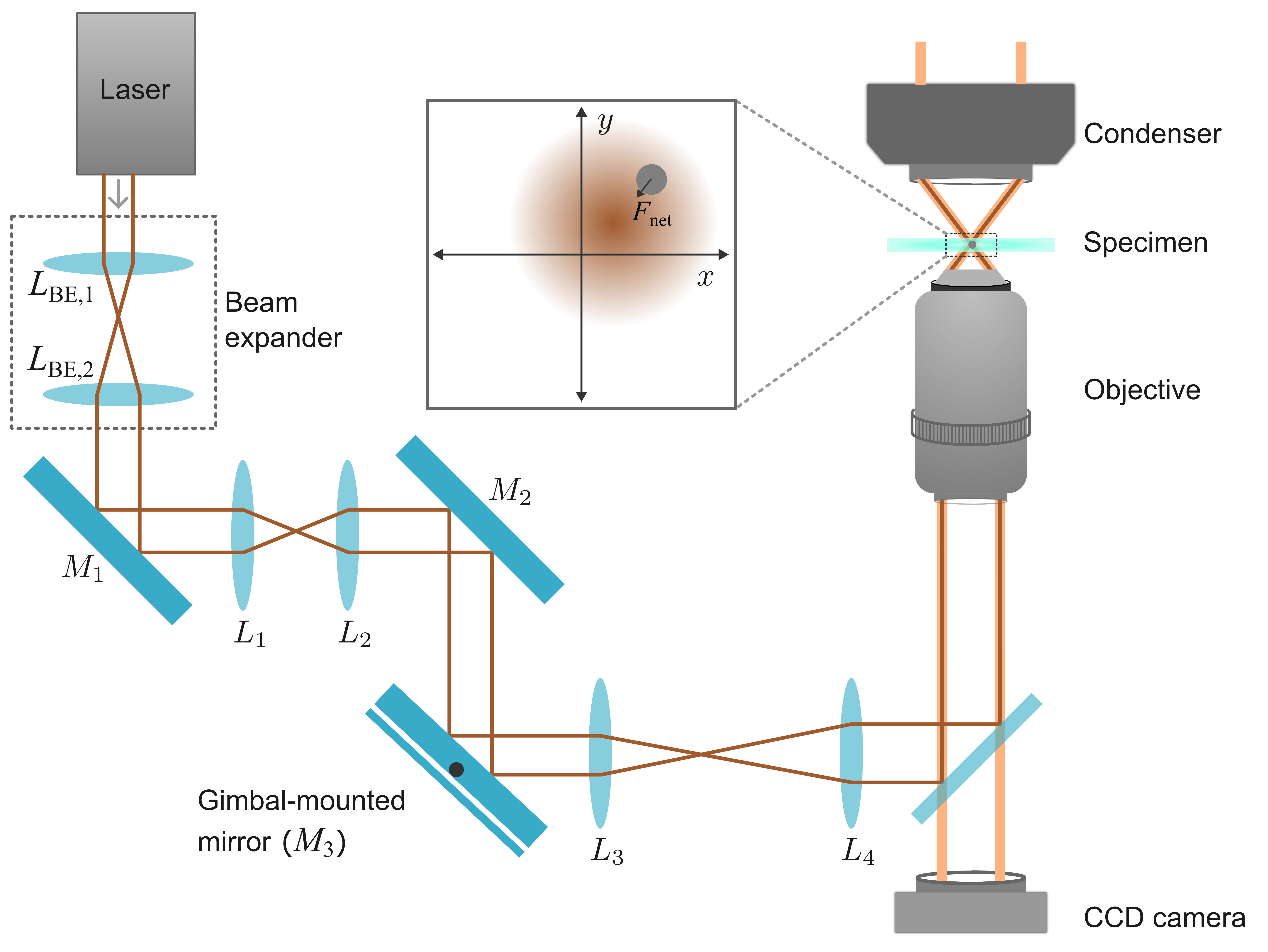"}
	\caption{Optical diagram of a single trap experimental setup. The component names have been abbreviated as: $L_{\textrm{BE,1}}$, $L_{\textrm{BE,2}}$, $L_1$, $L_2$, $L_3$, $L_4$ - lenses, $M_1$, $M_2$ - plane mirrors, $M_3$ - gimbal-mounted mirror. Brown lines depict the ray path of the laser. The enlarged view of the specimen plane shows a microsphere in an optical trap formed by a laser beam with $TEM_{00}$ mode; the particle is slightly displaced from the center of the trap, thus a net force ($F_{\textrm{net}}$) acts on the particle pulling it towards the trap center.}
	\label{fig:OT}
\end{figure*}

Starting with a two counter-propagating laser beam trapping setup, where the scattering force of the opposing beams is balanced at the middle where a micron-sized particle gets trapped \cite{Ashkin1970}, it was soon realized that the use of a single laser beam is more convenient \cite{Ashkin1986}. As previously discussed, a single laser beam can trap a dielectric particle when the gradient force is strengthened by tightly focusing the laser beam. A $TEM_{00}$ laser beam, which has a nearly Gaussian lateral intensity profile and does not change as the laser propagates, is the most popular choice for setting up an optical trap because it can be focused tighter and has a lower divergence. However, there are a few design considerations that are important to follow to set up an optical trap that is fully steerable in three dimensions \cite{Sheetz1998a, Khan2014}. The optical layout of such an optical trap is shown in Fig. \ref{fig:OT}, and the importance of each optical elements is discussed in the following paragraphs.

First, the laser beam is collimated and appropriately expanded using a beam expander. A beam expander (BE) is essentially a telescopic arrangement composed of a pair of biconvex or plano-convex lenses of focal lengths $f_{\textrm{BE},1}$ and $f_{\textrm{BE},2}$ placed at a separation of $(f_{\textrm{BE},1} + f_{\textrm{BE},2})$. The telescopic arrangement only changes the beam diameter by a ratio of $f_{\textrm{BE},2}/f_{\textrm{BE},1}$ without affecting the convergence of the beam. However, the inherent divergence of the laser beam can be corrected to obtain a perfectly collimated beam by appropriately increasing the separation between lenses. In principle, the expanded beam can be tightly focused through a high numerical aperture (N.A.) objective to form a functional optical trap. N.A. is deﬁned as the half-angle $\theta$ of the widest cone of light that can enter or exit the lens system. For a microscopic objective, N.A. is defined as N.A. $= n_{1} sin \theta$, where $n_{1}$ is the refractive index of the medium. A more convergent beam (larger $\theta$) produced by a high N.A. objective is used for more efficient trapping in advanced applications. This is because the gradient force is stronger when the light rays are bent to a greater degree; in other words, the peripheral rays contribute more effectively to the trapping efficiency compared to rays at the central part of the beam. Hence, oil-immersion objectives with N.A. as high as 1.4 or 1.45 are used for advanced and most effective optical tweezer setups. To utilize a high N.A. objective to its fullest potential, the entrance aperture (the back aperture, through which the laser beam enters) of the objective should be completely ﬁlled up by the laser beam. For this reason, the laser beam was first expanded. 

However, in practical applications, the trap must be steered in three dimensions, that is, both in the focal plane ($X-Y$) and along the optic axis ($Z$). We must use other optical elements to steer the trap freely in three dimensions without affecting its strength. To create a movable trap that provides the same trapping strength irrespective of its position ($X$ and $Y$) in the field of view, two important conditions must be fulfilled.
\begin{enumerate}
	\item The laser beam pivots around the centre of the entrance aperture of the microscope objective; and
	\item The laser beam must retain the same degree of filling of the entrance aperture when it pivots.
\end{enumerate}

The second condition is conveniently satisfied if the laser beam is expanded slightly wider such that it overfills the entrance aperture of the microscope objective. We will later see how this will also take care of the $Z$-movement of the trap with unaffected trapping strength. 

\begin{figure*}[ht] 
	\centering
	\includegraphics[width=1\linewidth]{"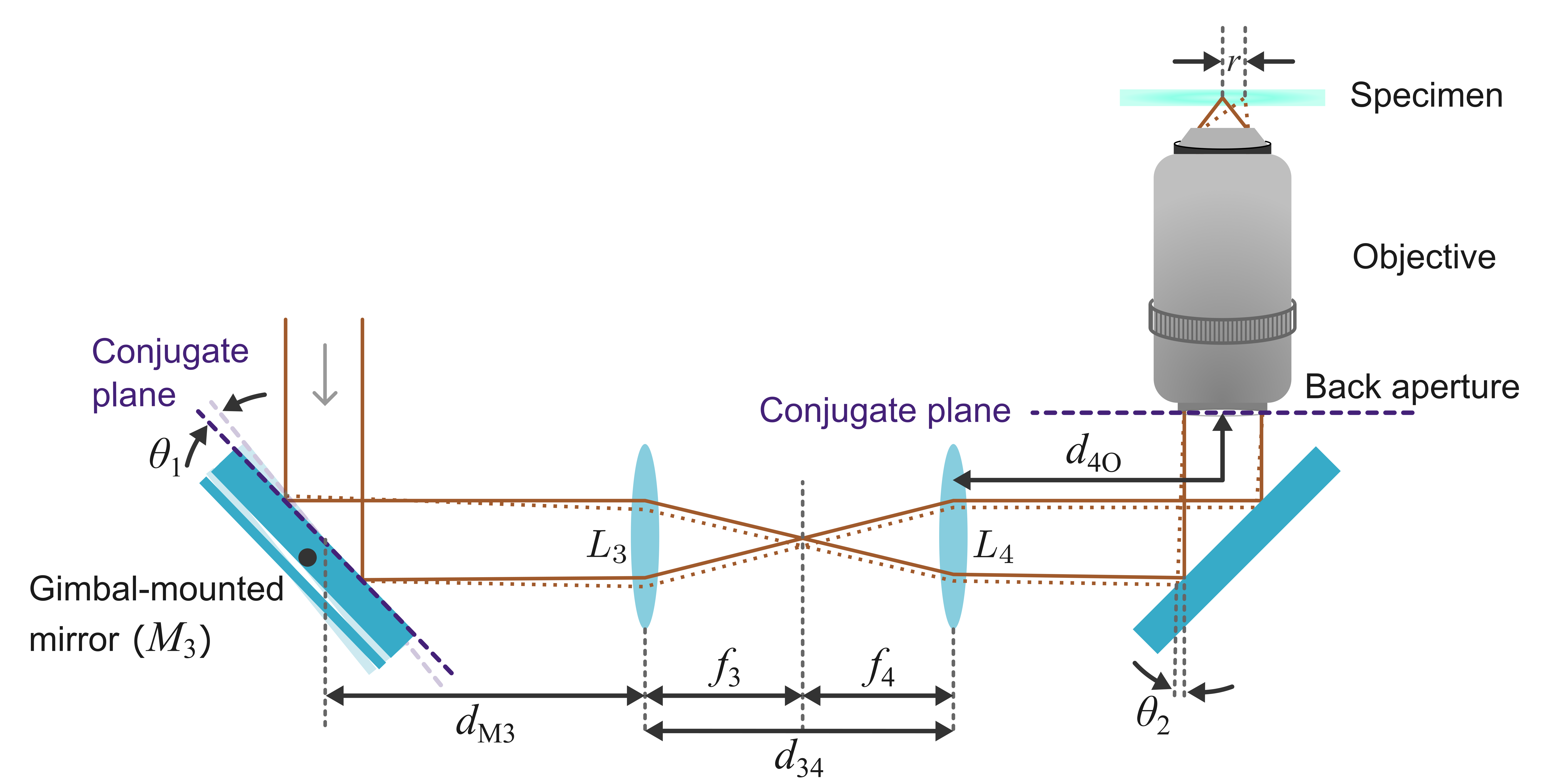"}
	\caption{The schematic depicts how an optical trap is steered in the focal plane ($X-Y$) by tilting a gimbal-mounted mirror ($M_3$). This is achieved by placing two lenses $L_3$ and $L_4$ in such a way that $M_3$ is at the conjugate plane of the objective back aperture. In this configuration, if $M_3$ is rotated by an angle $\theta_1$, the laser beam will enter the objective at an angle $\theta_2$, shifting the laser focus by a distance $r$ in the focal plane. Solid and dotted brown lines show the laser beam at two different orientations of the gimbal-mounted mirror.}
	\label{fig:XYS}
\end{figure*}

To satisfy the first condition, a gimbal-mounted mirror must be placed in the conjugate plane of the back aperture of the objective. This is achieved by imaging the entrance aperture of the objective on a gimble-mounted mirror using a system of lenses. The gimbal-mounted mirror can then be tilted to achieve the desired rotation of the laser beam around the center of the mirror. Because the mirror is placed at the conjugate plane of the entrance aperture, a small tilting of the mirror around its nominal position, which is deﬁned as the position at which the laser beam enters the objective parallel to its optic axis, will cause the laser beam to pivot around the center of the entrance aperture without any lateral movement. This, in turn, results in the movement of the trap in the focal plane ($X-Y$), as shown in Fig. \ref{fig:XYS}. A slight overfilling of the back aperture by the laser beam ensures that it always completely fills the aperture when it pivots around the center.  

Typically, a telescopic arrangement of two lenses is placed between the objective and the gimbal-mounted mirror for this purpose, maintaining the convergence of the beam. Thus, positioning the optical trap at a desired point in the focal plane by tilting the gimbal-mounted mirror ensures that the trap will not shift along the optic axis, that is, to a diﬀerent $Z$-plane. The trapping strength also remains the same irrespective of its movement in the $X-Y$ plane, as the beam fully fills the entrance aperture of the objective at all smaller angles (with respect to the optic axis) of incidence.

To image the objective entrance aperture at the center of the gimbal-mounted mirror, that is, to make them conjugate planes, the two lenses between $L_3$ and $L_4$, with corresponding focal lengths $f_3$ and $f_4$, should be placed at specific positions. If we denote the distances between the gimbal-mounted mirror and $L_3$, $L_3$ and $L_4$, and the $L_4$ and the microscope objective as $d_{\textrm{M}3}$, $d_{34}$ and $d_{4\textrm{O}}$ respectively, using the thin lens formalism we can derive the distance $d_{\textrm{M}3}$ in terms of the other parameters, as
\begin{equation}
d_{\textrm{M}3} = f_3 \frac{f_4 d_{34} - d_{4\textrm{O}}(d_{34} - f_4)}{f_4 (d_{34} - f_3) - d_{4\textrm{O}}(d_{34} - f_3 - f_4)}.	
\end{equation}
In general, the lens system with $L_3$ and $L_4$ is afocal to maintain convergence of the beam; therefore, we can set $d_{34} = f_3 + f_4$. Under these conditions, the expression for $d_{\textrm{M}3}$ becomes
\begin{equation}
d_{\textrm{M}3} = \frac{f_3}{f_4} (f_3 + f_4 - \frac{f_3}{f_4} d_{4\textrm{O}}).	
\end{equation}
In the most trivial case, $d_{4\textrm{O}}$ is set as $f_4$ to make $d_{\textrm{M}3}$ equal to $f_3$. With this arrangement, if the tilt angle of the gimbal-mounted mirror is $\theta_{1}$, corresponding angular deviation of the laser beam at the entrance aperture of the objective, $\theta_{2}$, can be given by,
\begin{equation}
\theta_{2} = -2 \frac{f_3}{f_4}	\theta_{1}.
\end{equation}
The shift of the optical trap in the focal plane, that is, in the $X-Y$ plane, $r$, is thus connected to $\theta_{2}$ through the effective focal length of the objective, $f_{\textrm{O}}$, as,
\begin{equation}
	r = f_{\textrm{O}} \theta_{2} = -2 f_{\textrm{O}} \frac{f_3}{f_4}	\theta_{1}.
\end{equation}
Therefore, the efficiency of steering the trap position by tilting a gimbal-mounted mirror is proportional to the ratio $f_{3}/f_{4}$ and $f_{\textrm{O}}$, which becomes a constant with the choice of the objective. The ratio $f_{3}/f_{4}$ is chosen as a compromise between the efficiency of the ability to move the trap in the focal plane and the expansion of the beam, given by the reciprocal, $f_{4}/f_{3}$, to overfill the objective back aperture. 

In an alternative realization, the entrance aperture of the objective is imaged onto the ﬁrst lens of the two-lens afocal system, that is, $L_3$. With this arrangement, lens $L_3$ can be moved laterally (in a plane perpendicular to the optic axis) to achieve a corresponding movement of the beam at the back aperture of the objective, and hence, a shift of the optical trap in the $X-Y$ plane. This realization also ensures a constant degree of ﬁlling of the objective entrance aperture, as the beam pivots around its center for small movements of the lens. However, there are some constraints on the focal lengths $f_3$ and $f_4$ required for this arrangement to work faithfully.

\begin{figure*}[ht] 
	\centering
	\includegraphics[width=1\linewidth]{"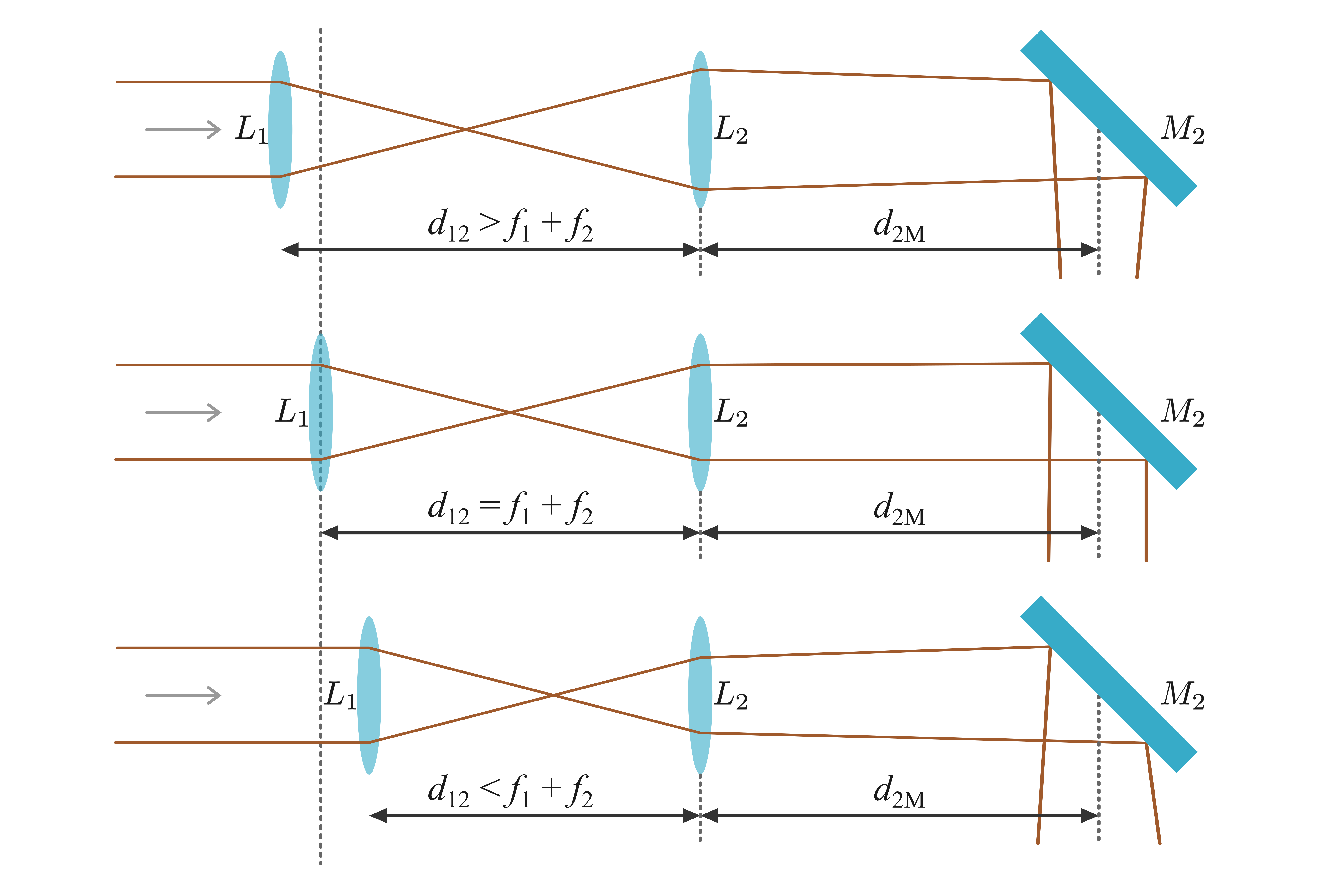"}
	\caption{The schematic demonstrates shifting the optical trap along the optic axis ($Z$-direction) by changing the convergence of the trapping beam. Here the beam becomes convergent to collimated to divergent, as the lens $L_1$ is shifted towards the right making the separation between the lenses $L_1$ and $L_2$ smaller in three steps, from top to bottom.}
	\label{fig:ZS}
\end{figure*}

To move the optical trap freely along the $Z$ direction with unchanged trapping strength, a second afocal optical system, formed by a pair of lenses $L_1$ and $L_2$ with corresponding focal lengths $f_1$ and $f_2$, is placed between the beam expander and the gimbal-mounted mirror. This can be realized with either two positive (biconvex or plano-convex) lenses, or one positive and one negative (concave) lens. To create an afocal optical system, the separation between the lenses, $d_{12}$, is nominally ﬁxed as the sum of the focal lengths of the two lenses, $f_1 + f_2$. By moving one of the lenses, say $L_1$, slightly along the optic axis, the otherwise collimated laser beam can be made slightly divergent or convergent, as depicted in Fig. \ref{fig:ZS}. This results in the beam focusing at a diﬀerent $Z$-plane as it emerges from the objective, thereby moving the trap at a different plane along the optic axis.

However, it is crucial to maintain the same degree of ﬁlling of the objective back aperture, as the beam is made convergent or divergent to move the trap along the $Z$-direction. The size of the laser beam around the center of the gimbal-mounted mirror reﬂects the degree of overﬁlling at the objective entrance aperture because the mirror is at the conjugate plane. Using thin lens formalism, the size of the laser beam at the gimble-mounted mirror, denoted as the beam diameter $D_{\textrm{M}}$, can be expressed in terms of the size of the incoming beam diameter, $D_{\textrm{in}}$, and the distances between the optical elements, that is $d_{2\textrm{M}}$, the distance between the gimbal-mounted mirror and $L_2$, and $d_{12}$, the distance between lenses $L_1$ and $L_2$. 
\begin{equation}
	D_{\textrm{M}} = - \left[ \frac{d_{2\textrm{M}}}{f_1} - \frac{(d_{12} - f_1)(d_{2\textrm{M}} - f_2)}{f_1 f_2} \right] D_{\textrm{in}}
\end{equation}   
From the above equation, it is evident that if $d_{2\textrm{M}}$ is set to $f_2$, the size of the laser beam at the gimbal-mounted mirror and therefore at the entrance aperture of the objective will be independent of $d_{12}$, that is, the separation between lenses $L_1$ and $L_2$. Therefore, $d_{12}$ can be varied appropriately to realize the desired convergence of the laser beam, and hence, the placement of the trap at the required $Z$-plane without affecting the degree of overﬁlling at the objective entrance aperture. The size of the beam at the gimbal-mounted mirror, in this realization, is given by,
\begin{equation}
	D_{\textrm{M}} = -  \frac{f_2}{f_1} D_{\textrm{in}}.
\end{equation}
This equation is valid regardless of whether the beam is collimated, convergent, or divergent. Note that a convergent beam (for $d_{12} > f_1 + f_2$) will focus at the lower Z-plane, whereas a divergent beam (for $d_{12} < f_1 + f_2$) will form a trap at a higher Z-plane. For a perfectly collimated beam (for $d_{12} = f_1 + f_2$), the focal plane of the trap coincides with the imaging plane of the objective. Therefore, with two sets of afocal arrangements, one before and one after the gimbal-mounted mirror, the resultant expansion of the beam can be estimated as,
\begin{equation}
	D_{\textrm{O}} =  \frac{f_2}{f_1} \frac{f_4}{f_3}D_{\textrm{in}},
\end{equation}
where $D_{\textrm{O}}$ denotes the beam diameter at the entrance aperture of the objective. Considering the expansion of the laser beam that is coming out of the laser head with diameter $D_{\textrm{L}}$ provided by the beam expander, the total expansion factor of the laser beam is calculated as,
\begin{equation}
	D_{\textrm{O}} =  \frac{f_{\textrm{BE},2}}{f_{\textrm{BE},1}} \frac{f_2}{f_1} \frac{f_4}{f_3}D_{\textrm{L}}.
\end{equation}

\subsubsection{Dual Optical Traps}

\begin{figure*}[ht] 
	\centering
	\includegraphics[width=1\linewidth]{"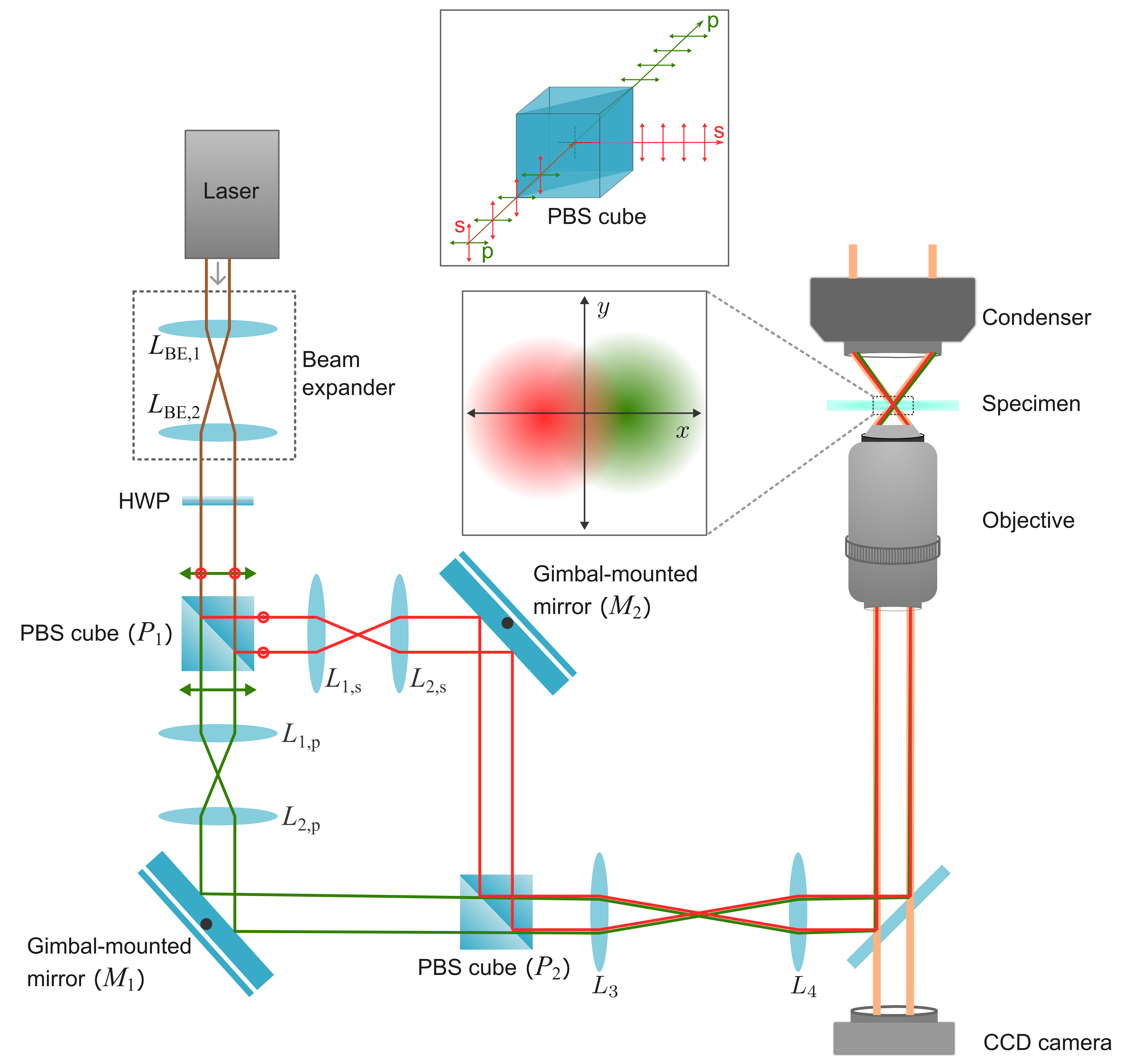"}
	\caption{Optical layout of a dual trap experimental setup. The component names have been abbreviated as: HWP - half-wave ($\lambda/2$) plate; $L_{\textrm{BE,1}}$, $L_{\textrm{BE,2}}$, $L_{\textrm{1,s}}$, $L_{\textrm{2,s}}$, $L_{\textrm{1,p}}$, $L_{\textrm{2,p}}$, $L_3$, $L_4$ - lenses, $M_1$, $M_2$ - gimbal-mounted mirrors, $P_1$, $P_2$ - polarizing beam splitter cubes. The red and green lines show the ray path of the s- and p-polarized components of the laser beam. The enlarged view of the specimen plane shows two optical traps in red and green originating from s- and p-polarized components, respectively. The traps are separated by some distance as the gimbal-mounted mirrors, $M_1$ and $M_2$, are not perfectly parallel, making a slight angle between them. \textit{Inset}: Working principle of a polarizing beam splitter (PBS) cube. It splits an incident light into two orthogonally polarized beams by allowing one polarization to pass undeflected (p) and reflecting the orthogonal polarization (s).}
	\label{fig:DOT}
\end{figure*}

In most practical applications, having two independently steerable traps side-by-side enhances the experimental capabilities by many folds \cite{Khan2011a}. Using a polarizing beam-splitter cube, a laser beam can be conveniently split into two orthogonally polarized beams (Fig. \ref{fig:DOT} Inset), each creating an independent trap at the specimen plane \cite{Khan2014}. This is particularly convenient because the two laser beams can be merged into a common optical path entering the objective using a second polarizing beam-splitter cube (which works as a beam-merging cube) with negligible loss in laser power. Most of the high-quality lasers that are commonly used for optical micromanipulation applications are linearly polarized, and the plane of polarization can be rotated in the desired direction using a half-wave ($\lambda/2$) plate. Thus, by rotating the $\lambda/2$ plate, which is placed just prior to the beam splitter cube, the plane of polarization of the incoming beam, and hence the relative intensities of the parallel and perpendicular polarizations that govern the splitting ratio, can be conveniently controlled. Moreover, the two trapping beams with orthogonal polarizations do not interfere with each other, and hence can be manipulated independently, as if they are from two different lasers.

Forming two traps from orthogonal polarizations will be beneficial only if the two traps can be steered independently in three dimensions. Hence, the optical elements that allow steering of the laser beam in three dimensions must be independent for these two beams with orthogonal polarizations, as shown in Fig. \ref{fig:DOT}. Because the polarizing beam splitter cube allows one polarization to pass undeflected and reflects the orthogonal polarization, the two beams move along perpendicular directions after being split by the cube. Two gimbal-mounted mirrors are used to bring the two beams back to the second polarizing beam splitter cube at a right angle. The second cube merges two beams to subsequently follow a common path. Moreover, the two gimbal-mounted mirrors, one for each beam, steer the beams independently, resulting in decoupled lateral movements (in $X-Y$) of the corresponding traps in their respective focal planes. It is important to note that the two beams of orthogonal polarizations create a small angle between them as they enter the objective; hence, they are focused at different points in the focal plane. The angle between the beams is tuned using the two gimbal-mounted mirrors to achieve the desired separation between the traps or the specific positioning of each of the traps in the $X-Y$ plane. The afocal lens arrangement formed by $L_1$ and $L_2$, which allows $Z$-movement of the trap, is also placed separately in the beam paths of the two beams, allowing independent vertical shifting of the traps. After the second beam splitter cube, the two beams pass through the same set of optics before entering the objective at a small relative angle.

Any optical micromanipulation technique remains incomplete, unless positional fluctuations of the trapped microsphere or other specimens can be detected. In a dual-trap setup, sensing the positions of both trapped objects in real time with a high spatial resolution is required. In a later section, we discuss the position detection techniques and their importance in practical applications.

\subsection{Contemporary Techniques}

There have been many improvisations on traditional single Gaussian beam optical tweezers, and other completely new tweezer conﬁgurations have been developed in recent years. In this section, we discuss the improvements and developments. Each of these techniques has speciﬁc applications and comes with its own limitations. However, the working principles remain the same - utilizing the photon momentum of a laser beam for manipulation–that is, to confine, push, or drag micron- and submicron-sized dielectric materials in a suspension.

\subsubsection{Tweezers Based on Novel Light Modes}

Non-Gaussian modes other than the $TEM_{00}$ mode, which is commonly used in conventional optical tweezers setups, are also employed to trap and manipulate dielectric particles \cite{He1995,He1995a,Bains1996,Friese1996,Simpson1996,Simpson1997,ONeil2001,Grier2003,Meyrath2005,Lafong2006,Milne2007,Jia2008,Meresman2009,Ambrosio2010}. Higher-order laser beams, such as the Hermite Gaussian beam ($TEM_{xy}$), Laguerre-Gaussian (LG) beams ($TEM_{pl}$), and Bessel beams are typical examples. Optical tweezers based on LG beams offer unique possibilities and advantages over the conventional Gaussian beams. The most important advantages include better trapping efficiency along the axial ($Z$) direction and the capability to trap particles with lower refractive index contrast or with higher reflectance and absorbance.  An LG beam also possesses a well-deﬁned orbital angular momentum that can apply torque and thus rotate a trapped particle without any external mechanical or electrical steering. Bessel beams are advantageous because they satisfy the Helmholtz equation and thus are propagation invariant, as they retain their cross-sectional intensity profile upon propagation, unlike a Gaussian beam. Bessel beams have concentric rings, where the zeroth order has a bright spot at the center and the higher orders possess a dark region at the center. Thus, a long array of traps with lateral translation, which acts as an optical conveyer belt transporting submicron particles over longer distances, can be formed by interference patterns from two Bessel beams.

\subsubsection{Time-shared Multiple Traps}

Multiple optical traps or customized optical potentials are often required for many advanced applications such as molecular motor studies. The most convenient way to create multiple traps is by time-sharing a single beam with an acousto-optic deflector (AOD) or using a piezo-driven scanning mirror that can deflect the beam in the desired direction \cite{Guilford2004,Bustamante2021}. These elements usually replace the gimble-mounted mirror and thus sit at the conjugate plane of the objective back-entrance aperture. The AOD or piezo driven mirror time shares the laser beam between each site with an appropriate residence time, or scans over a desired path to create a customized potential pattern. A slower diffusing micron-sized object feels a time-averaged potential averaged over its self-diffusion timescale. Therefore, high-frequency switching of the beam from one site to another, or continuous scanning over a pattern, effectively creates a static trap pattern. One-dimensional line traps, circular traps, and arrays of optical traps are commonly created using this technique. Here, the limitation comes from the scanning or switching rate of the beam-deflecting device.  

\subsubsection{Holographic optical tweezers}

Another convenient way of creating intricate trap patterns, even in three dimensions, is by using specially designed diﬀractive optical elements, specifically holographic techniques or phase-contrast methods \cite{Dufresne2001,Schmitz2005,Grier2006,Mejean2009}. The phase and intensity profiles of the laser beam are modulated by a reflecting or transmitting spatial light modulator (SLM). When the modulated beam is focused through the objective, the phase profile creates an intensity profile, which is the Fourier transform of the phase profile. An SLM is an effective dynamic hologram made from an array of liquid crystals that can induce a phase change from 0 to 2$\pi$ at the trapping wavelength. Therefore, the SLM can be readily programmed for appropriate phase and intensity modulations to create a trap pattern with the desired intensity and phase profile. Dynamic holograms created on SLM are employed with various algorithms, and can create a large number of traps, where even the mode structure of each trap can be speciﬁed individually, arrays of optical vortices, or holographic line traps that can be placed at any desired position in three dimensions. Nucleation seeds with trap patterns matching the lattice structure for larger colloidal crystals or templates for advanced studies of cell growth can also be formed.

\subsubsection{Optical Fiber Based Traps}

If the end of an optical ﬁber tip moulded into a lens-like facet, it focuses the light to form a steep-gradient optical trap at the desired precise location. Thus, a single-mode optical ﬁber can create an optical trap in 3D without using any external optics \cite{Constable1993,Lyons1995, Sidick1997, Collins1999,Renn1999,Guck2001}. However, if the shape of the tip of the ﬁber is not adequate to tightly focus the light beam, a counter-propagating beam from another fiber at the opposite end is required to balance the scattering force. This technique of two counter-propagating beams from two fibers has also been used to stretch microparticles, where the stretching force can be optimized by appropriately tuning the beam intensities. In an optical ﬁber-based setup, multiple traps with different mode structures can be formed using multimode optical fibers. Intricate excitation of diﬀerent optical modes of the ﬁber is required to realize different trapping geometries. Cells and clusters of cells have been oriented and rotated under a microscope using optical fiber-based traps \cite{Kreysing2008}. The main advantage of this technique over conventional optical tweezers is access to the precise trapping location and thus less distortion and loss of laser power, and decoupling of trapping from imaging optics.

\subsubsection{Near-field Optical Trapping}

Although conventional optical trapping is performed in the far-field region, operating in the near-field offers some advantages. The evanescent field, which is a residue optical ﬁeld that ‘leaks’ during total internal reﬂection and fades oﬀ at an exponential rate, is used to manipulate a microbead. Hence, this technique has a significantly reduced trapping region within approximately 100 nm from the surface \cite{Kawata1992,Clapp1999,Lester1999,Clapp2001,Lutti2007}. A continuous evanescent ﬁeld with directional sense is produced owing to multiple internal reflections when a light beam propagates through an optical waveguide. Thus, microparticles can be guided along the propagation direction of the light beam. A more recent version of evanescent ﬁeld optical tweezers employs extended optical landscapes to propel a large number of particles along a preferred direction without the use of waveguides. This is termed ‘lensless optical trapping’ (LOT). 

Electromagnetically coupled pairs of metallic nanostructures illuminated by a laser beam produce a localized evanescent field; thus, an array of subwavelength plasmonic optical traps can be created to trap nanometer-sized particles suspended in water or air \cite{Novotny1997,Grigorenko2008, Righini2008, Yang2009}. Another realization involves the use of an evanescent field near a nanoaperture or a nanostructured substrate. Although the working principle of these techniques is exciting and important, they are difficult to implement in practice. Moreover, the nanostructure-induced optical near-ﬁelds reduce the trapping volume beyond the diﬀraction limit. However, if implemented, nanometric tweezers have the potential to provide nanoscale control of entities at signiﬁcantly lower laser powers, allowing the nano-manipulation of fragile biological macromolecules.

\section{Capabilities of Optical Micromanipulations}

\subsection{Application and Measurement of Force}

The ability to manipulate micron- and sub-micron-sized particles by applying precise forces at the pico-Newton level makes optical tweezers a unique technique. Many new optical tool-kits, such as optical spanners \cite{Simpson1996,Simpson1997} and holographic optical tweezers \cite{Zemanek1998}, have been developed over the years to widen the applicability. Here, we discuss how an optical trap can be employed to apply both scattering and gradient forces to a trapped particle and to a system to which the bead is attached. Moreover, optical tweezers are used to conveniently measure an external force with sub-piconewton to femtoNewton resolution. However, proper calibration is required to either apply predetermined forces or measure externally applied forces. 

\subsubsection{Application of Optical Force}

Two types of optical forces, the gradient force and scattering force, can be applied through a focused laser beam. Micron- and sub-micron-sized dielectric particles are manipulated using these forces in an appropriate fashion. Bulk media can also be perturbed on a micrometer length scale through a trapped bead. It is important to note that many design and system parameters must be tuned properly to maximize the effect of the applied force \cite{Sheetz1998a,Khan2014}.

The radiation pressure of the laser beam exerts a scattering force along the direction of beam propagation. Thus, it is nonconservative in nature and can propel micron-sized objects downstream \cite{Ashkin1970}. To effectively push micron or submicron particles and transport them from one point to another using the scattering force, the gradient force that attracts them towards the focal spot must be reduced. This can be achieved in many ways, such as not focusing the beam tightly, removing the peripheral rays that contribute more to the gradient force, or slightly defocusing the beam as it enters the objective. The same radiation pressure can induce a torque on micro-objects that have ﬁnite chirality. A chiral object can transform a linear force to a torque by virtue of its geometry, and thereby rotates about the optical axis, where the gradient force provides a hinge to the rotating particle \cite{Galajda2001,Khan2005,Khan2006,Khan2007,Khan2009}. A birefringent particle can also rotate and apply torque when trapped by circularly polarized light \cite{Bradac2018}. 

The gradient force is a restoring force; hence, it pulls a particle to a mechanical equilibrium at the point of focus. Thus, it can hold a micron- or submicron-sized object nonmechanically at the focal spot. Even moving micro-objects, such as motile bacteria or spermatozoa, can be held static by a gradient force.  In the opposite scenario, a moving trap can pull a trapped bead and thereby apply an effective force on the system to which the bead is attached, for example, a cell membrane. Similarly, a ﬂowing medium drags along a microbead  until it becomes trapped in the gradient force ﬁeld. In this case, the gradient force not only seizes the motion of the microbead but also applies shear to the ﬂowing medium by creating a relative local displacement field. This constitutes a very effective method to apply shear in active microrheology experiments. However, this process changes significantly depending on the density, viscosity, or local microstructures in the medium and stiffness of the optical trap.

\subsubsection{Measuring External Forces}

\begin{figure*}[ht] 
	\centering
	\includegraphics[width=1\linewidth]{"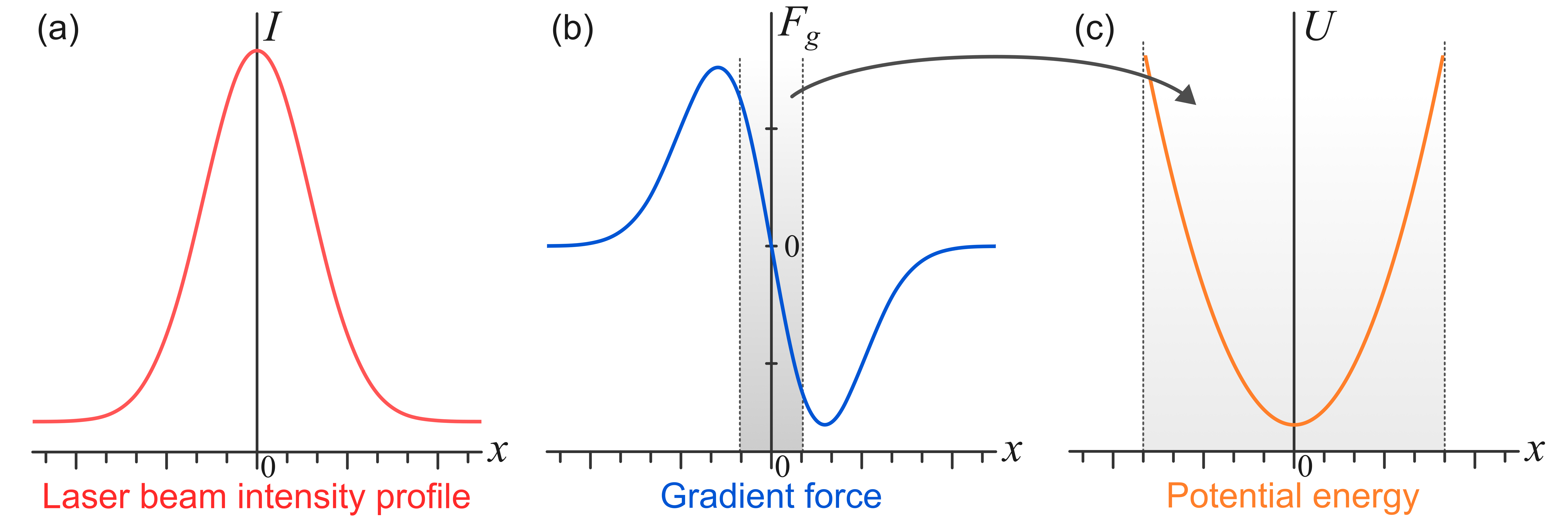"}
	\caption{The gradient force and associated potential corresponding to a Gaussian beam profile. (a) A tightly focused beam with Gaussian intensity distribution generates a gradient force that is linear near the center (b), and hence acts as a restoring force. (c) The potential landscape corresponding to the linear region is parabolic in shape.}
	\label{fig:RF}
\end{figure*}

The gradient force generated by a Gaussian ($TEM_{00}$) laser beam is a linear restoring force ($F_x= \kappa_x \Delta x$), where $\kappa_x$ is the force constant along the $x$ direction) for small displacements ($\Delta x$) from the center, similar to the force applied by a simple spring in a mechanical system \cite{Sheetz1998a,Khan2014}. The gradient of the Gaussian intensity proﬁle (Fig. \ref{fig:RF} (a)) is shown in Fig \ref{fig:RF} (b). The potential landscape corresponding to the linear region is parabolic ($\kappa_x x^2$) in shape (Fig \ref{fig:RF} (c)) \cite{Khan2011a}. Thus, an optical trap not only applies force to the dielectric particle but can also measure the applied force on the trapped bead from its displacement from the center. The stiffness, that is the force constant of an optical trap must be determined prior to the measurement of external forces using different calibration techniques, which will be discussed later. In the case of a properly aligned optical trap with a symmetric beam profile, the lateral force constants $\kappa_x$ and $\kappa_y$ mostly have very close values. However, the axial confinement is comparatively much weaker because of the lower intensity gradient along the axial direction, and the associated force constant ($\kappa_z$) is significantly smaller.  When working with small particles, the eﬀect of polarization becomes significant, and the $\kappa_x$ and $\kappa_y$ values may differ considerably.

Under the influence of an external force (other than the trapping forces), the trapped bead is displaced from the center of the trap, which is otherwise the equilibrium position. At the displaced position, $x_{\textrm{d}}$, the gradient force balances the $x$-component of the external force, $F_{\textrm{ext},x}$, and a new mechanical equilibrium is established, where,
\begin{equation}
	F_{\textrm{ext},x} = \kappa_x x_{\textrm{d}}.
	\label{ForceBalance}
\end{equation}
A similar equation can be written with $y_{\textrm{d}}$ and $\kappa_y$ if the external force has a nonzero $y$ component. The displaced position of the trapped particle is usually tracked using a position-sensing diode or quadrant photodiode. Particle tracking from a recorded video can also provide the lateral displacement of the bead from the center of the trap, $x_{\textrm{d}}$, and $x_{\textrm{d}}$. The external force is then precisely calculated using Eq. \ref{ForceBalance}  and the value of the force constant, $\kappa$.

Using this technique, external forces, such as the drag force exerted by a ﬂowing medium on a microbead or the force required to stretch a coiled molecule that is attached to the trapped bead and pulled from the other end, can be measured with high precision. Appropriate tuning of the force constant (by adjusting the laser beam power) to maximize the displacement of the trapped bead, which depends on the magnitude of the external force, enhances the sensitivity and reliability of the measurement.

\subsubsection{Calibration of Force Constant}

The stiffness or the force constant ($\kappa$) of an optical trap arises from the interaction between the dielectric particle in the trap and the electromagnetic field of the focused laser beam near the focal spot. Thus, its value depends on many parameters associated with the interaction, such as particle size, shape, refractive index, beam geometry, and laser power. However, in practice, the force constant is conveniently adjusted by changing the laser power if the properties of the trapped particles remain the same, which is often the case for a set of similar experiments. Therefore, trap stiffness is often calibrated against the laser power for a given experimental setting using the same particle and medium. The value of $\kappa$ ($\kappa_x$ and $\kappa_y$) can be determined using a number of methods, each with its own advantages. Here, we discuss the three most commonly used techniques: the equipartition, power spectrum, and drag force methods \cite{Sheetz1998a,Khan2014}. 

\begin{figure*}[ht] 
	\centering
	\includegraphics[width=1\linewidth]{"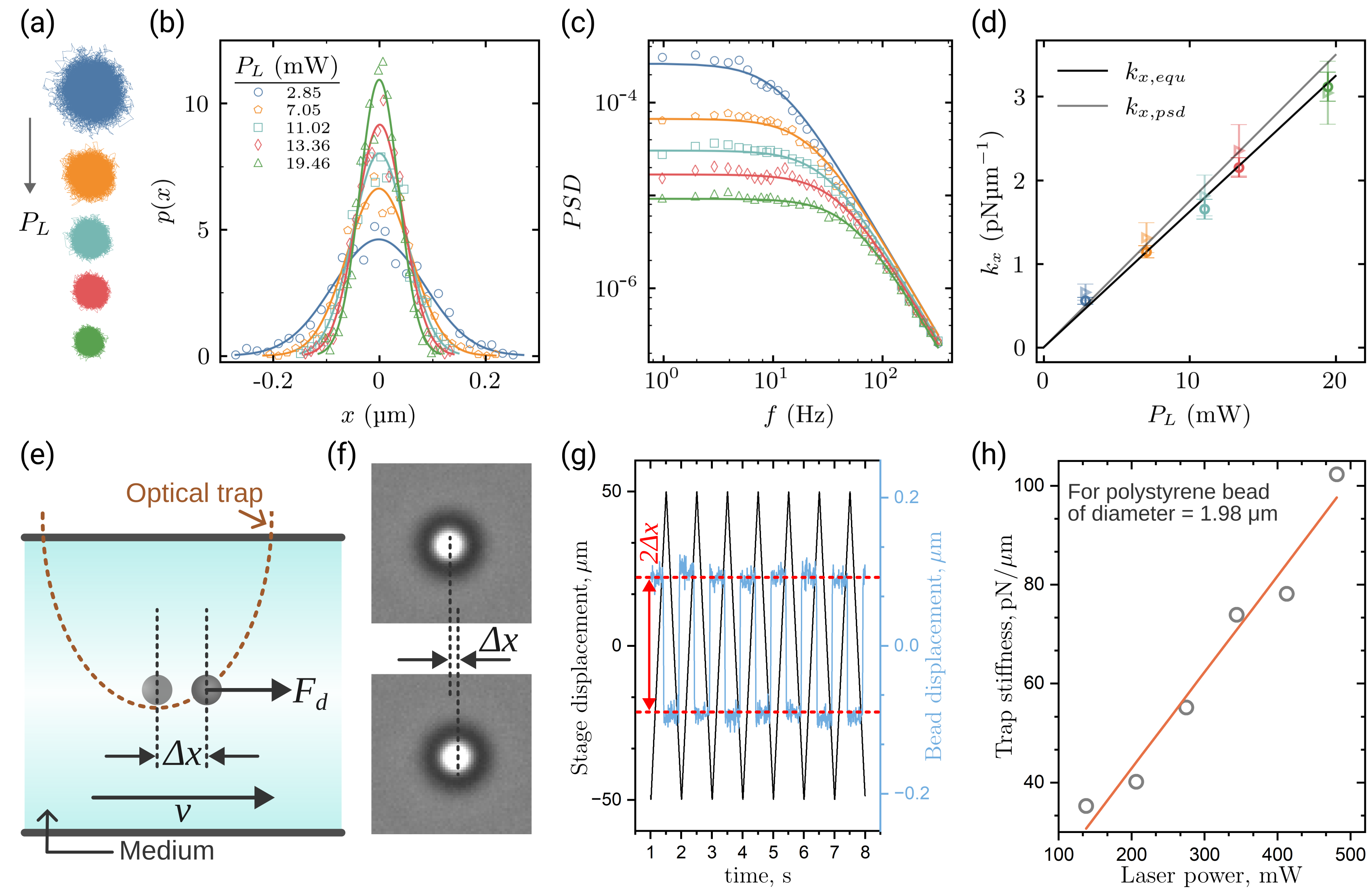"}
	\caption{Different methods for trap stiffness calibration. The confined trajectories of a microbead are captured at increasing trapping laser intensity (a) to obtain the positional probability distribution along $X$ (b) or $Y$. The fitting of the positional probability distributions along $X$, $p(x)$, with Gaussian distributions provides the variance ($\left\langle x^2\right\rangle $) for each $p(x)$. Moreover, good agreement with the Gaussian fits demonstrates the absence of artifacts. (C) The corresponding single sided power spectral densities are fitted with Lorentzian, and show good agreement. The linear relation of the force constant $\kappa_x$ (along $X$) with laser power $P_L$, measured using equipartition and power spectral density methods are shown in (d). Schematic (e), and microscope images (f) show the displacement ($\Delta x$) of a trapped microsphere in response to the drag force $F_{\textrm{d}}$ applied by translating the sample stage at a constant speed $v$, inducing the same relative motion to the medium in reference to the trapped bead. The measured variation in the displacement of the bead and stage is shown in (g). The trap stiffness determined using the drag method too increases linearly with the laser power (h).}
	\label{fig:CAL}
\end{figure*}

To determine the trap stiffness using the equipartition method, the equilibrium positional fluctuations of a trapped particle due to spontaneous thermal excitations are recorded to calculate the average potential energy along $X$ and $Y$. According to the equipartition theorem for a particle confined in a harmonic potential \cite{Sheetz1998a,Khan2014},
\begin{align}
\frac{1}{2} \kappa_i <p_i^2> &= \frac{1}{2} k_{\textrm{B}}T,  \; \; \; i = x, y \\
\textrm{or,} \; \; \; \; \; \; \; \; \; \; \; \; \;  \kappa_i  &= \frac{k_{\textrm{B}}T}{<p_i^2>}, \; \; \; i = x, y
\end{align}
where $p_i$ denotes the generalized position measured along coordinates $i$ that is $x$ and $y$, and $k_{\textrm{B}}$ and $T$ are the Boltzmann constant and temperature, respectively.  Thus, the positional variance of a trapped particle provides trap stiffness using the equipartition theorem. In addition to its simplicity, the principal advantage of this method is that it does not explicitly depend on other parameters such as the size and shape of the particle or the viscosity of the surrounding medium. Nonetheless, in fairness, all the properties, such as the size and shape of the particle and the optical and viscous properties of the medium, will influence the trap stiffness through the positional ﬂuctuations of the particle. The most crucial part of this method is to reliably track the position of the trapped particle. Because variance is an intrinsically biased estimator (it is derived from the square of a quantity and is therefore always positive), any added noise and drift in position measurements results in an increase in the overall variance, thereby decreasing the apparent stiffness estimate. However, the presence of these artifacts becomes apparent by the deviation of the positional probability distribution from a Gaussian distribution, which is expected for a harmonically bound Brownian particle. Fig. \ref{fig:CAL} (a) and (b) show the typical position ﬂuctuations of a trapped particle, which follows a Gaussian distribution. The trap stiffness is calculated from the variance of the probability distribution determined by the fitting parameter for a Gaussian fit. In Fig. \ref{fig:CAL} (d), the linear increase in the trap stiffness along $x$ ($\kappa_{x, \textrm{equ}}$), measured using the equipartition method, is plotted against the laser power ($P_L$).

When a microbead of known (Stokes) radius ($a$) is trapped in a medium of known viscosity ($\eta$), the Langevin equation for a harmonically bound Brownian particle can be used to determine the force constant ($\kappa$) of the optical trap \cite{Sheetz1998a,Khan2014}. The Langevin equation (in one dimension for simplicity) is given as,
\begin{equation}
	m \ddot{x}(t) + \gamma \dot{x}(t) + \kappa_x x(t) = f(t).
\end{equation}
Here $x(t)$ is the trajectory of the trapped Brownian particle of mass $m$, and friction coefficient $\gamma$ ($= 6\pi \eta a$). The RHS, $f(t)$, represents the random Brownian forces. In the Fourier space, that is as a function of the Fourier frequency $f$, and defining $\kappa_x = 2 \pi \gamma f_0$, this equation can be given by,
\begin{equation}
	\left| X(f) \right| ^2 4 \pi^2 (f_0^2 + f^2) = \frac{1}{\gamma^2} 	\left| F \left\lbrace f(t)\right\rbrace  \right| ^2.
	\label{Spectral}
\end{equation}
By definition $	\left| X(f) \right| ^2 \equiv S_x$   and $\left| F \left\lbrace f(t)\right\rbrace  \right| ^2 \equiv S_f$ are the single-sided power spectral densities of $x(t)$, and $f(t)$, respectively.  By applying the ﬂuctuation dissipation theorem, we obtain $S_f  = 4 \gamma k_{\textrm{B}}T$. Therefore, Eq. \ref{Spectral} simplifies to,
\begin{equation}
	S_x = \frac{k_{\textrm{B}}T}{\pi^2 \gamma} \frac{1}{f_0^2 + f^2}.
\end{equation}
The functional form of $S_x$ describes a Lorentzian function with corner frequency $f_0$ and a plateau value $S_0$, occurring at $f \to 0$. Thus, fitting the single-sided power spectral density of the positional fluctuation, $x(t)$, of a trapped bead with a Lorentzian provides the value of $f_0$, which in turn determines the value of $\kappa_x$ as $\kappa_x = 2 \pi \gamma f_0$ \cite{Khan2010}. Fig. \ref{fig:CAL} (c) shows a few typical variations in $S_x$ at a monotonically varying laser power. The linear increase in the trap stiffness ($\kappa_{x, \textrm{psd}}$) obtained from the Lorentzian fits to $S_x$ with increasing laser power ($P_L$) is shown in Fig. \ref{fig:CAL} (d).

A trapped microbead can be displaced from its equilibrium position in the trap by applying a constant external force of known magnitude, and the trap stiffness can be determined simply by measuring the displacement of the bead in the most direct manner \cite{Sheetz1998a,Khan2014}. A constant force of known magnitude can be conveniently applied by creating a relative streamlined flow of constant speed ($v_x$) along the $X$ direction with respect to the trapped bead. Knowing the viscosity of the fluid ($\eta$) and Stokes radius of the bead ($a$), the drag coefficient ($\beta$) and the drag force ($F_{\textrm{d}}$) are calculated as $\beta = 6 \pi \eta a$, and $F_{\textrm{d}} = \beta v_x$, respectively. The relative flow is usually generated by translating the sample stage at a constant speed ($v_x$) in a back and forth motion for repeated measurements, while the bead remains trapped at a displaced position ($\Delta x$), along the same direction, from the center of the trap. However, because the Stokes drag force that displaces the particle arises from the hydrodynamics around the trapped bead, the drag coefficient, including any surface proximity corrections, must be known. A triangular displacement of amplitude $A_x$ and frequency $f$ of the sample stage along the $X$ direction results in a square-wave profile of the relative velocity, and hence, the force, and the corresponding displacement of the trapped microbead along the same $X$ direction. For each period of motion, the bead trajectory can be given by,
\begin{equation}
	\Delta x (t) = \frac{\beta A_x f}{2 \kappa_x}\left[ 1 - exp \left( - \frac{\kappa_x}{\beta} t \right) \right]  ,
\end{equation}
where $\kappa_x$ is the trap stiffness along the $X$ direction. This equation also includes Faxen’s correction because of the proximity of static surfaces. When the particle is trapped at a height ($\sim 50 a$) where the Faxen’s law correction is negligible, this equation reduces to the simple Stokes drag equation,
\begin{equation}
	\Delta x (t) = \frac{\beta}{\kappa_x} v_x (t) = \frac{6 \pi \eta a v_x (t)}{\kappa_x}.
\end{equation}
Thus, following the drag method, the trap stiffness along $X$ and $Y$ can be determined as,
\begin{equation}
\kappa_{x}  = \frac{6 \pi \eta a v_x (t)}{\Delta x (t)}, \;\; \textrm{and} \;\; \kappa_{y}  = \frac{6 \pi \eta a v_y (t)}{\Delta y (t)}.
\end{equation}
Fig. \ref{fig:CAL} (e) shows the schematic of the experiment, and Fig \ref{fig:CAL} (f) and (g) exhibit a typical displacement of the position of the bead, as observed using video microscopy and measured using a quadrant photo diode for the corresponding triangular stage displacement profile, respectively. A linear relationship between trap stiffness and trapping laser power is shown in Fig. \ref{fig:CAL} (h).

\begin{figure*}[ht] 
	\centering
	\includegraphics[width=1\linewidth]{"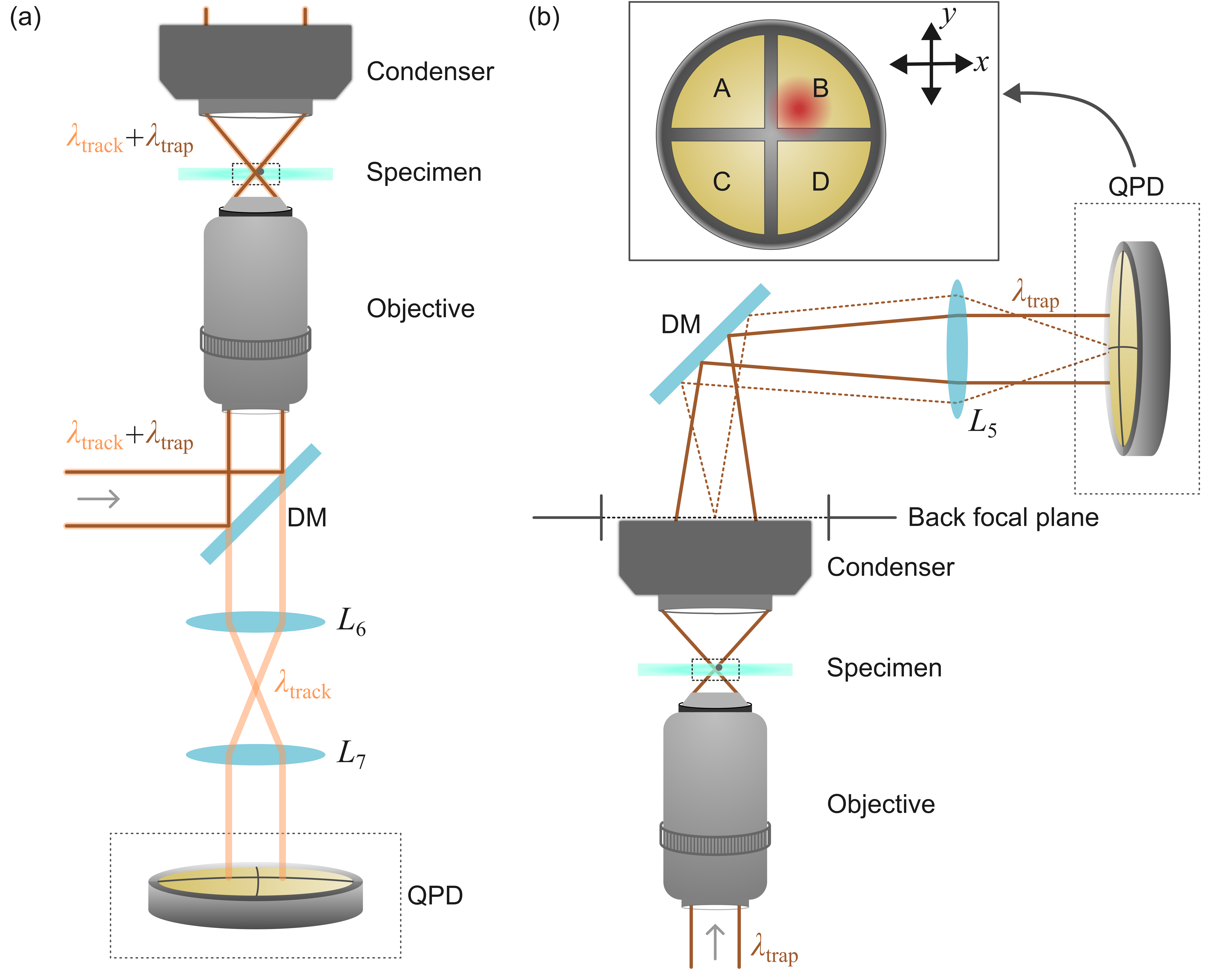"}
	\caption{Tracking the position of the trapped particle using QPD. (a) A separate tracking laser ($\lambda_{\textrm{track}}$), collinear to the trapping laser ($\lambda_{\textrm{trap}}$), is focused on the trapped particle, and the back scattered light is captured through the objective. The back aperture of the objective is imaged on the QPD to track the particle from the deviation of the back scattered light. (b) The trapping laser ($\lambda_{\textrm{trap}}$) becomes forward-scattered by the trapped particle and is collected by the condenser. The deviation of the forward-scattered light is captured by the QPD, which is placed at the conjugate plane of the back aperture of the condenser. The inset shows how the collimated light falls asymmetrically on the QPD and generates an output signal as the particle moves away from the center of the trap.}
	\label{fig:PT}
\end{figure*}

For all trap stiffness calibration techniques, or in general, for all optical trap-based experiments and applications, the most crucial part is to record the instantaneous position of the trapped particle with finer resolution and at a high rate. The time series of the position coordinates of the trapped bead, $x(t)$ and $y(t)$, can be tracked in two broad ways, using video imaging or with a position sensing module, such as a quadrant photo diode (QPD) or position sensing diode (PSD). Although it is easier to integrate with the setup, the video imaging technique has two shortcomings: a relatively low bandwidth (typically at fps of 30 - 500) and critical dependence on the particle detection and tracking algorithm. In contrast, in the indirect method, the deviation of the forward or backscattered light from the trapped bead can be detected using a QPD or PSD to obtain the translational movement of the particle with high precision (sub-nanometer) and at a higher bandwidth ($\sim$ 20 kHz). This requires an additional set of optical elements to image the deviation of the scattered light onto the sensor, that is, the QPD. After passing through the trapped particle, the forward-scattered trapping laser is collected and collimated by the condenser. The translational movement of the trapped particle in the focal plane causes the collimated forward-scattered light to pivot around the center of the back aperture of the condenser. A lens arrangement is used to image the back aperture of the condenser on to the QPD, as shown in Fig. \ref{fig:PT} (b) \cite{Gittes1998}. Thus, even the slightest movement of the trapped particle in reference to the center of the trap results in an amplified deviation of the collimated beam on the QPD surface. The QPD signal is processed by a differential amplifier, which provides a linear variation in the output voltage as a measure of the translational movement of the bead. In an alternate arrangement, a separate collinear tracking laser is sent to the sample along with the trapping laser \cite{Khan2014}. Because of the low power and small beam waist, the tracking laser does not generate a significant gradient force to add to the trapping efficiency. The tracking laser is collected and collimated by the objective itself after being backscattered by the trapped particle. In this case, the back aperture of the objective is imaged on the QPD using a lens arrangement (Fig. \ref{fig:PT} (a)). A similar deviation of the backscattered tracking laser on the surface of the QPD provides positional fluctuation of the trapped bead. While aligning the tracking laser, making it collinear with the trapping laser, is nontrivial, this method makes the tracking process and its efficiency independent of the trapping strength or the positioning of the trap. In contrast, position detection using the forward-scattered trapping laser provides the displacement of the bead in reference to the center of the trap; hence, it becomes a more convenient choice when working with a moving or scanning optical trap.

\subsection{Localized Heating}

When exposed to incident light, a scatterer in the form of a single atom, molecule, liquid, or solid radiates a part of the incident electromagnetic energy through its accelerated charge carriers. Apart from the reradiation of the energy, the charge carriers also absorb part of the incident energy to get thermally excited, raising the local temperature \cite{Sheetz1998a, Bustamante2021}. Thus, the absorption of the trapping laser in the optical trap causes local heating. Although local heating due to absorption of the trapping laser light may cause damage to the sample that is in the trap or in the proximity of the trap, it also opens up new applications. To minimize harm to biological samples, near-infrared lasers are mostly used in optical tweezers for biological applications because biological materials absorb less at longer wavelengths ($\lambda >$ 800 nm), and water absorption, which peaks far-infrared (3 $\mu$m), is only moderate at near-IR wavelengths \cite{Svoboda1994}.  

The absorption of light not only causes local heating but also induces a non-conservative force, $ F_{\textrm{A}}$, that is similar and complementary to the scattering force ($F_{\textrm{S}}$) \cite{Bohren1998}. The resultant nonconservative force field ($F_{\textrm{NC}}$) can be given by,
\begin{equation}
F_{\textrm{NC}} = F_{\textrm{S}} + F_{\textrm{A}} = \frac{n (\sigma_{\textrm{S}} + \sigma_{\textrm{A}})}{c}\langle \vec{S}_i \rangle
\end{equation}
where $n$ is the refractive index of the medium, $\sigma_{\textrm{S}}$ and $\sigma_{\textrm{A}}$ are the scattering and absorption cross-sections, respectively, and $\langle \vec{S}_i \rangle$ is the time- averaged Poynting vector. This extra force, along with local heating, becomes significant for particles with asymmetric absorption cross sections, such as Janus colloids, where one hemisphere of the spherical particles is metal coated, bringing in a wide range of experimental possibilities, from the thermophoretic active propulsion of Janus colloids to the measurement of local temperature in fluids.

Metal-coated Janus or metal nanoparticles-embedded colloids generate a local temperature gradient around them owing to absorption when placed in an optical trap. In the case of a half-coated Janus colloidal particle, the absorption is dominant at the metal-coated sides (for a 10 nm gold thin film, absorption $\sim$ 30 \% at $\lambda$ = 492 nm) \cite{Loebich1972,Yakubovsky2018}, creating an asymmetric local temperature gradient around the particle. The Janus colloid experienced thermophoretic self-propulsion because of the asymmetric temperature gradient. While the self-propulsion direction is solely determined by the Soret coefficient ($S_T$), the speed ($v$) depends on the temperature gradient ($\nabla T$ ) as well, and is given by, $v = - D S_T \nabla T$, where $D$ is the diffusivity of the particle \cite{Jiang2010,Buttinoni2012}. This thermophoretic self-propulsion provides an excellent system for conveniently studying free or confined active Brownian dynamics under a defocused or tightly focused optical trap, respectively. Local asymmetric heating of carbon-capped Janus particles under laser radiation also causes dissociation of the critical mixture on the hotter side, resulting in diffusiophoretic self-propulsion of the particles \cite{Lavergne2019}. In contrast, metal-nanoparticles-embedded colloids generate a circularly symmetric temperature gradient along the radial direction. The azimuthally symmetric radial thermophoretic force induces an effective confinement of particles, resulting in optothermal trapping \cite{Paul2022}. 

The temperature field at the microscopic length scale can also be mapped using optical tweezers-based fluorescent thermometry. For this, a temperature-sensitive fluorescent dye (Rhodamine B, Rhodamine 6G, etc.) is excited by a focused laser beam, usually an optical trap, working at a wavelength within the excitation window, and its emission, whose intensity also depends on the local temperature, is imaged to map the temperature of the fluid \cite{Ross2001,Duhr2004,Natrajan2008,Zhou2019}. Although the quantum efficiency, and hence the emission of these dyes, changes with temperature, their absorption depends on several factors, including the pH of the medium, which is also dependent on the local temperature.  

Many biological processes are modulated by temperature, such as enzymatic reaction rates and opening/closing of membrane channels \cite{Clapham2003}. The heterogeneity in temperature might also have a regulatory effect on intracellular biomolecules and their functions \cite{Okabe2012}. Thus, these phenomena can be regulated by controlling the local temperature using an optical trap. The use of gold nanoparticles to enhance local heating provides an effective solution for many applications, including photo-thermal cancer therapy \cite{Gobin2007}, contraction of cardiomyocytes \cite{Oyama2012}, and vesicle transport \cite{Oyama2012a}. This has been applied even at the single-molecule level, such as in the activation of a single myosin. The stepping rate of myosin V, which is attached to gold nanoparticles, was accelerated by a few folds, while the maximum force decreased to almost half of that of ones attached to polystyrene beads.

\subsection{Cutting and Ablation}

Different laser wavelengths (UV or IR) with varied energy per photon and the desired laser power of a focused laser microbeam are used to employ them as scalpels or scissors, allowing precise cutting with sub-micrometer resolution \cite{Colombelli2004}. For the first time, in 1969, an argon ion laser was used to create half-micron-diameter lesions in individual chromosomes \cite{Berns1969a,Berns1969}. Subsequently, an increased energy density in the focused spot of the argon laser microbeam could alter chromosomes in live cells \cite{Berns1971,Basehoar1973}. Later, laser micromanipulation was performed independently of imaging, using dichroic mirrors \cite{Berns1971a,Berns1974}. This simple and versatile laser microbeam was used in many similar studies over a decade until pulsed lasers and different dye lasers became available for application as laser scissors \cite{Berns1981}. The ability of a focused laser beam to cut through or fuse a cell membrane or tissue, in its application as scissors, and for ablation of organelles has been utilized in many different forms based on the requirements. These wide range of applications include genetic surgery \cite{Berns1969,Berns1969a}, chromosome movements during mitosis or meiosis \cite{McNeill1981}, and DNA damage recognition and repair \cite{Wang2013,Truong2013, Wang2014}. These capabilities become enhanced manifolds when multiple microbeams are combined to work simultaneously as laser tweezers to hold, as scalpel and scissors to dissect, and for ablation.  

A focused laser beam for trapping and ablation is also useful in other systems, such as ablative  photodecomposition in solid materials for surface modification and fabrication of polymers \cite{Srinivasan1983, Srinivasan1989}, as well as for microelectronic applications \cite{Dupuy1989}. Both the high-energy state photochemical reaction and locally heated thermal reaction induced by high peak power  and short pulse width lasers play crucial roles in the ablation of solid surfaces. A laser trapping-spectroscopy-ablation system was employed by coupling a CW and a pulsed Nd$^{3+}$: YAG laser with an optical microscope  to achieve laser ablation along with simultaneous fluorescence  spectroscopy with nanosecond time resolution  on an optically trapped PMMA latex  particle in water \cite{Misawa1990}. Three-dimensional optical trapping and laser ablation of a single polymer latex particle in water). Laser ablation of an  optically trapped  microcapsule containing  pyrene/toluene in water and optical micromanipulation of small pyrene/toluene oil droplets ejected  from the capsule have also been demonstrated \cite{Misawa1991}.

\section{Optical Micromanipulations of Soft Materials for Devices and Technologies}

\subsection{Characterization of Viscoelastic Properties}

The viscoelastic properties of soft materials, that is, their complex mechanical response to external perturbations, are arguably the most pertinent characteristics for determining their usability in devices and technologies. These properties have been conventionally studied in industries employing (bulk) rheology, where the entire sample is sheared and the stress response is measured. A more recent technique, invented in 1995, called microrheology, revealed that it is possible to probe the mechanical response of soft materials, even at microscopic length scales \cite{Mason1995,Mason1997}. Since then, it has been widely used to probe the local viscoelastic properties of materials, where heterogeneity plays an important role, and to study scale-dependent rheological properties. Although microrheology was invented to extract the viscoelastic properties of the environment from the trajectories of a probe microbead, optical trapping greatly enhances its capabilities. Optical micromanipulation offers a powerful platform for performing microrheological measurements from single molecules to near-macroscopic scales in both linear and nonlinear regimes. The principal advantage of optical tweezers-based microrheology measurements is that they allow measurements over a wider frequency range, and active shear can be applied to push the system to a nonlinear regime employing active-microrheology techniques \cite{Addas2004,Bishop2004,Squires2005,Wei2008,Khan2010,Wilson2011,Neckernuss2015,Furst2017,RobertsonAnderson2018,Khan2019}.

The working principle of the more widely used passive microrheology is to extract the frequency-dependent elastic modulus $(G^{\prime}(\omega))$ and viscous modulus $(G^{\prime\prime}(\omega))$ of soft materials from the thermal diffusion of an embedded microsphere through the generalized Stokes-Einstein relation \cite{Mason1995,Mason1997, Squires2010,Furst2017}. In an improvised optical-tweezer-based version, the positional fluctuations of a trapped microbead conveniently provide the viscoelastic response ($G^{\prime}(\omega)$ and $G^{\prime\prime}(\omega)$) of the surrounding fluid, relying on the fluctuation-dissipation theorem (FDT), over a wider frequency range. The FDT enables us to derive the viscous part ($\chi^{\prime\prime}(f)$) of the response function of the system, $\chi(f) = \chi^{\prime}(f) + i \chi^{\prime\prime}(f)$, directly from the equilibrium positional ﬂuctuation of the trapped bead, $x(t)$ (or $y(t)$), through the following equation \cite{Addas2004,Atakhorrami2006,Khan2010}:
 \begin{equation}
 	\chi^{\prime\prime}(f) = \frac{\pi}{2 k_{\textrm{B}} T}f S(f),
 \end{equation} 
where $k_{\textrm{B}}$, $T$, and $S(f)$ are the Boltzmann constant, temperature, and single-sided Power Spectral Density (PSD) of the positional time series $x(t)$, respectively. The real part, ($\chi^{\prime}(f)$), representing the elastic response, is dependent on the viscous response ($\chi^{\prime\prime}(f)$) through the Kramers-Kronig relation, and can be derived as,
 \begin{equation}
	\chi^{\prime\prime}(f) = \frac{2}{\pi} \int_{0}^{\infty} dt \; cos(ft)  \int_{0}^{\infty} d\xi \; sin(\xi t) \chi^{\prime\prime}(\xi)
\end{equation} 
The elastic response of the system is overestimated because the optical trap response is elastic in nature. To subtract the trap effect, the response function is corrected as,
\begin{equation}
	\alpha(f) = \frac{\chi(f)}{1 - \kappa \chi(f)} 
\end{equation} 
where $\alpha(f)$ is the corrected system response function and $\kappa$ is the trap stiffness. The complex shear modulus of the system, $G(f) = G^{\prime}(f) + i G^{\prime\prime}(f)$, is related to the complex response function through,
\begin{equation}
	G(f) = \frac{1}{6 \pi a \alpha(f) } ,
\end{equation} 
where $a$ denotes the Stokes radius of the bead. Therefore, the storage and loss moduli can easily be calculated in terms of the response functions as,
\begin{align}
	G^{\prime}(f) &= \frac{1}{6 \pi a } \frac{\chi^{\prime}}{ \chi^{\prime 2} + \chi^{\prime\prime 2}} - \frac{\kappa}{6 \pi a}, \\
	G^{\prime\prime}(f) &= \frac{1}{6 \pi a } \frac{\chi^{\prime\prime}}{ \chi^{\prime 2} + \chi^{\prime\prime 2}}.
\end{align}

Active microrheology is a more widely employed technique because it provides enhanced capabilities in terms of applying an active shear to the system. An optically trapped microsphere is often driven through a soft material, and the mechanical response to strain is measured from the displacement of the same trapped bead with respect to the center of the trap. Both the instantaneous velocity of the sample stage ($v_x$) and the displacement of the trapped bead ($\Delta x$) are measured at high spatial resolution and high bandwidth to obtain precise microrheological measurements. The strain rate ($\dot{\gamma_x}$) is measured from the velocity ($v_x$) as $\dot{\gamma_x} = 6v_x/a$, where the prefactor ($=6$ here) depends on the boundary condition \cite{Squires2007}, $a$ is the Stokes radius of the trapped bead, and the stress response ($\sigma_x$) is obtained from $\sigma_x = \kappa \Delta x$, where $\kappa$ is the stiffness of the trap. The strain rate-dependent viscosity ($\sigma(\dot{\gamma})$) has been conveniently measured for various soft materials that are more liquid-like using this simple method \cite{Squires2005,Khan2010,RobertsonAnderson2018}. 

\begin{figure*}[ht] 
	\centering
	\includegraphics[width=1\linewidth]{"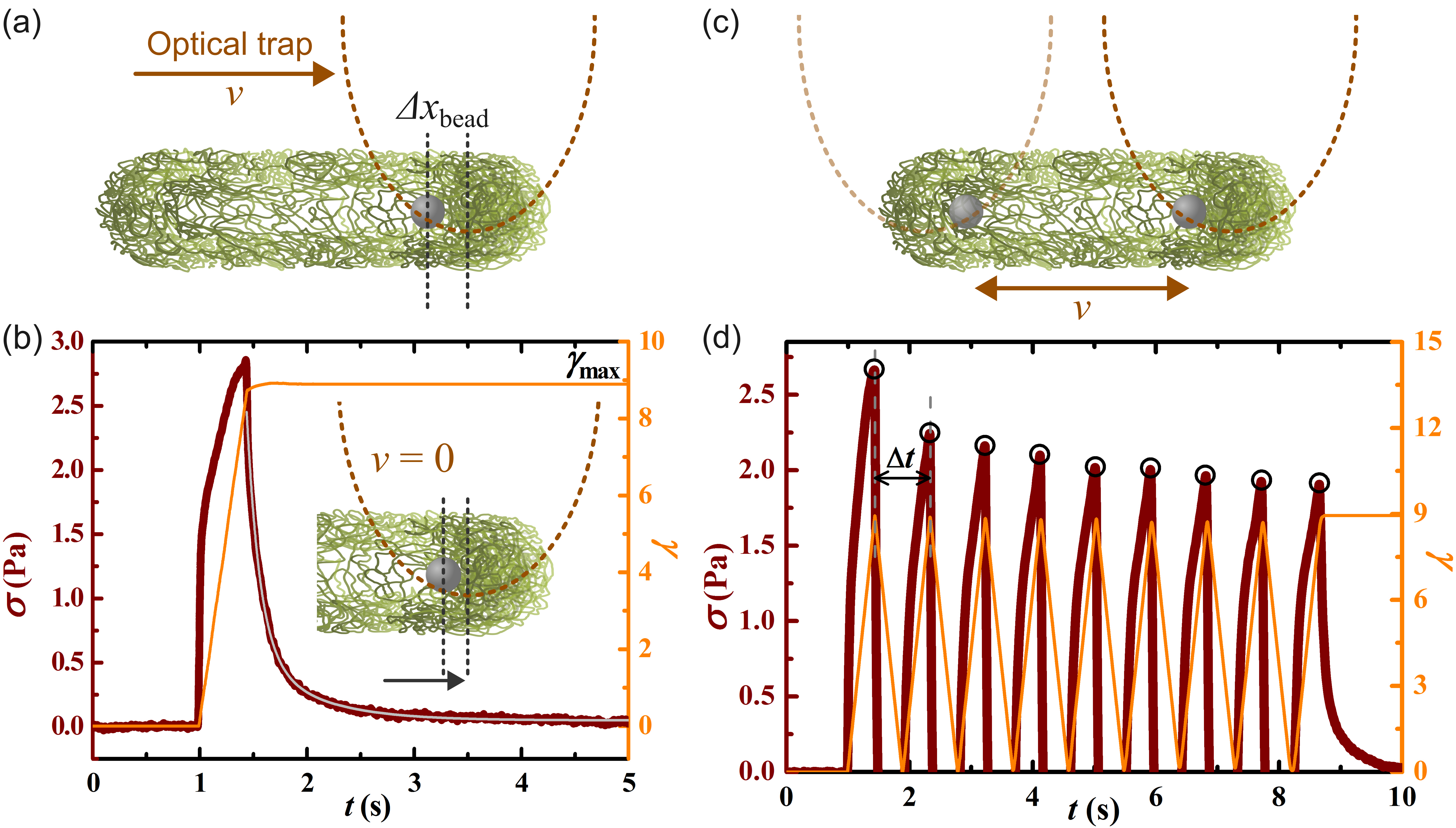"}
	\caption{Active microrheology of entangled biopolymer (DNA) network. (a) A microsphere was dragged through the network suspension by applying active shear to the system. The speed of the bead ($v$) provides the shear rate ($\dot{\gamma}$), or in turn, the shear strain  ($\gamma$), and the instantaneous displacement of the bead from the center of the trap ($\Delta x_{\textrm{bead}}$) measures the shear stress ($\sigma$). When the translational motion is stopped, the displaced bead returns to the center of the trap, exhibiting the stress relaxation behavior of the sample, as shown in the inset. (b) The stress buildup while the probe particle applies shear to the system, and subsequent stress relaxation after cessation of shear, are plotted (brown) using the left axis, while the strain profile (orange) is shown using the right axis. (c) In oscillatory shear measurement, the trapped bead is dragged to and fro with a sinusoidally varying or constant speed ($v$). (d) Typical time-varying stress ($\sigma$) and strain ($\gamma$) are plotted in brown and orange, respectively, using the left and right axes \cite{Khan2019}.}
	\label{fig:AMR}
\end{figure*}

A trapped microsphere is also driven through a soft material with a high relative speed to apply a large shear rate, thereby pushing the system to a strongly non-linear regime. When the relative movement suddenly stops, the system is left to relax. The stress relaxation of the system is recorded by the decreasing distance between the position of the trapped bead and the center of the trap, $\Delta x (t)$. The stress relaxation curve, $\sigma(t)$ reveals the relaxation times ($\tau$) and respective relaxation mechanisms of the system \cite{Helfer2000,RobertsonAnderson2018,Khan2019}. In an alternate measurement protocol, the optical trap is switched off with cessation of the drive. The microspheres exhibit recoil in response to the elastic recovery of the material as soon as it is set free by switching off the trap. These recoil trajectories also reveal the elastic stress recovery of soft materials \cite{Chapman2014,RobertsonAnderson2018,Ginot2022}.  

The microrheological analogs of the oscillatory shear experiments are implemented by driving an optically trapped bead in a small-amplitude back-and-forth motion, following a sinusoidal strain rate profile ($\gamma (t)$) with a specific frequency ($\omega$). The instantaneous displacement of the trapped bead from the center of the trap, $\Delta x(t)$, measures the time-varying stress profile, $\sigma(t)$. Their relative amplitude and phase difference ($\Delta \phi$) provide the storage modulus, $G^{\prime} (\omega)$, and loss modulus, $G^{\prime \prime} (\omega)$ of the system at $\omega$ through,
\begin{align}
	G^{\prime} (\omega) &= [\sigma_{\textrm{max}} / \gamma_{\textrm{max}}] cos \Delta \phi, \\
	G^{\prime \prime} (\omega) &= [\sigma_{\textrm{max}} / \gamma_{\textrm{max}}] sin \Delta \phi, \textrm{and} \\
	\eta^* (\omega) &= [(G^{\prime} (\omega))^2 + (G^{\prime \prime} (\omega))^2 ]^{1/2} / \omega.
	\end{align}

The measurement is repeated at varying frequencies ($\omega$) to derive the frequency-dependent moduli, $G^{\prime} (\omega)$, $G^{\prime \prime} (\omega)$ \cite{Chapman2014a,Neckernuss2015,RobertsonAnderson2018,Knezevic2021}. Large-amplitude and high-strain-rate oscillatory shear measurements have also been performed using optical tweezer-based active microrheology \cite{Khan2019}. 

Optical-tweezer-based microrheology is a contemporary and evolving field. Many new measurement protocols are applied, and existing protocols are improvised to address the specific requirements of the study of viscoelasticity in soft materials. Some examples include dual oscillating trap active microrheology experiments \cite{Paul2018} and simultaneous measurements of active and passive microrheology \cite{Khan2010}.

\subsection{Dynamics Force Spectroscopy}

Measuring and sensing dynamic changes in small forces, down to the order of a few picoNewtons, in response to microscopic elongations or displacements, commonly called dynamic force spectroscopy, is a fundamental part of making devices for technologies based on biomaterials. Although other techniques are available for force spectroscopy, such as magnetic tweezers and atomic force microscopy (AFM), optical tweezers are arguably the most versatile and convenient single-molecule manipulation tool. It enables us to exert forces on the order of a few pN to an excess of 100 pN on particles ranging in size from nanometers to micrometers and simultaneously detect its displacements in three dimensions with sub-nanometer accuracy at sub-millisecond time resolution. Herein, we discuss the different applications of dynamic force spectroscopy using optical tweezers.  

Optical tweezers-based dynamic force spectroscopy is widely used to pull out a membrane tether, which is a lipid nanotube from the cell surface, to measure the mechanical properties of the cell membrane and the underlying receptor–cytoskeleton linkages \cite{Baoukina2012}. While a constant or linearly increasing force is applied by pulling a trapped bead attached to the cell membrane, a non-linear transition in the force-distance plot signifies the formation of a membrane tether \cite{Dai1995}. The tether force and membrane viscosity can be determined from the slope of the force-distance plot at the transition \cite{Hochmuth1996}. For example, it was revealed that membrane tension is a continuum property over the entire cell surface, and the force required to pull a tether is determined by membrane–cytoskeleton linkage \cite{Dai1999}. More recent studies that combined fluorescence imaging with an optical trap demonstrated that actin exists in pulled membrane tethers \cite{Pontes2011}. Free membrane-like behavior showing a movable tether–membrane junction has also been observed in neuronal axons \cite{Datar2015}.

\begin{figure*}[ht] 
	\centering
	\includegraphics[width=1\linewidth]{"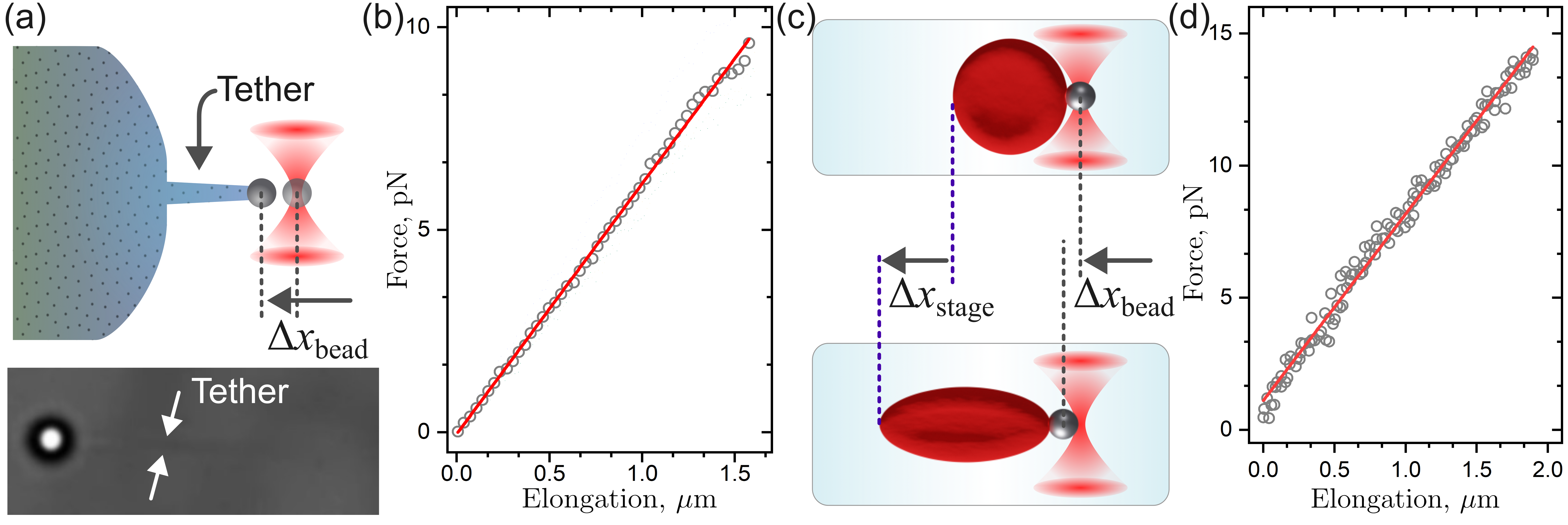"}
	\caption{Dynamic force spectroscopy on cells by measuring the force required for stretching at different elongation rates. (a) A tether is formed when a microbead attached to the cell membrane is pulled by a translating optical trap. The tether is a thin tube-like structure that remains invisible or poorly visible under bright-field microscopy. (b) The force-elongation plot provides the stiffness of the tether, which often varies with the pulling rate and elongation length. (c) In some cases, such as for human erythrocytes, the cell as a whole becomes elongated when a bead attached to its membrane is pulled by translating either the trap or the stage, that is, the substrate where one end of the cell is anchored. (d) A similar force-elongation plot is obtained, providing the elasticity of the cell, which contains elastic contributions from both the cell membrane and cytoskeleton.}
	\label{fig:FS}
\end{figure*}

Some cells, such as human erythrocytes (red blood cells or RBCs), become fully deformed when a membrane-attached microbead is pulled using optical tweezers \cite{Svoboda1994,Dao2003}. Mechanical properties, such as shear modulus and relaxation time, were studied by providing uniaxial deformations of varying amounts to RBCs. To this end, RBCs are elongated either by fixing one end of the cell to the cover glass and pulling it with a bead attached at the opposite end or by stretching it with two optical handles attached at diametrically opposite positions \cite{Henon1999,Dao2003,Lim2004,Dao2006, Yoon2008,Yoon2011}. Employing an alternate approach, three time-shared optical traps were used to fold a trapped RBC, mimicking the deformation of an erythrocyte in narrow capillaries, to discover younger cells that have shorter relaxation times compared to the population \cite{Bronkhorst1995}.

Dynamic force spectroscopy on malaria-infected red blood cells provides a great tool to distinguish healthy RBCs from pathologically infected RBCs by comparing their shear moduli. The Pf155/Ring-infected erythrocyte surface antigen (RESA), introduced by \textit{Plasmodium falciparum} merozoites, significantly reduced the deformability of infected RBCs and even more so at febrile temperatures \cite{Mills2007}. The adhesive strength between merozoites and RBCs was also measured \cite{Crick2014,Introini2022}, and Dantu RBCs, a blood group that protects against malaria, were found to have higher average tension, providing greater resistance against parasite invasion \cite{Kariuki2020}. Weak optical traps have also been used to observe membrane fluctuations in red blood cells, allowing an indirect measurement of membrane bendability and stiffness \cite{Betz2009}. Recent membrane fluctuation studies have exhibited a violation of the fluctuation-dissipation relation, demonstrating the non-equilibrium active nature of healthy RBC membrane flickering. Other notable studies on erythrocyte deformability using optical tweezers have been discussed in these review articles \cite{Suresh2006,Kim2015,Zhu2020,Avsievich2020,Xie2021,Matthews2022}.

In another set of dynamic force spectroscopic applications, the stepping characteristics of highly dynamic motors in protein systems have been studied. This has helped us understand the fundamental aspects of how various motor proteins, such as myosin, kinesin, and dynein, move on the microtubules and actin ﬁlaments and, more importantly, how the cargo, such as molecules, droplets, and organelles, are transported by the motors along the cytoskeletal tracks over a long distance. Moreover, from a future application perspective, it is more important to discern how ATP hydrolysis fuels the mechanical work done by motors \cite{Sweeney2018}. The pivotal roles and contributions of optical tweezers-based techniques in advancing this field over the last two decades have been well described in these review articles \cite{Molloy2002, Moffitt2008, Choudhary2019,Wang2022}. The two-step working stroke of a myosin-I head attached to an actin filament and the associated biochemical states of the actomyosin cycle were first discovered in 1999 by achieving a time resolution of ~ 1 ms for the measurements \cite{Veigel1999}. Further improvements in the time resolution allowed the characterization of the two mechanical states of a fast myosin-V motor \cite{Uemura2004}. It has been shown that working stroke for myosin decreases with the load, and the forward motion of kinesin-14 HSET can be stalled with a load of 1.1 pN \cite{Reinemann2018}. Although the motility of motors is stalled if the load is increased above a certain value, a fluctuating force causes acceleration in kinesin motors \cite{Ariga2021}. 

Optical tweezers have also been used to study the force spectroscopy associated with the mechanical unfolding of proteins and nucleic acid structures \cite{Kellermayer1997,Liphardt2001}. The force measurement capabilities of optical traps are ideally suited for nucleic acid folding, which  corresponds to a $\sim$15 pN force for $\sim$ 1 nm displacements. Using a birefringent particle in an optical trap made of circularly polarized light \cite{Friese1998,LaPorta2004,Deufel2007}, or a chiral particle in a linearly polarized trap (see section \ref{sec:microrotor}) can apply torque and induce rotation. The torque required to buckle DNA as a function of the applied load has been measured using this technique \cite{Deufel2007}.

\subsection{Mechanosensitivity}

In many biological systems, the binding energies of molecular bonds and the opening of ion channels are governed by mechanical stress or other mechanical stimuli. Optical tweezers have become a convenient and popular tool to measure the mechanosensitivity, that is the response of these systems to a mechanical stress or strain applied by the optical trap; moreover, to  examine  how extracellular mechanical stimuli such as pressure, stretching, and motion are transformed into intracellular biochemical signals via a number of mechanoreceptors on the cell membrane \cite{Chen2017}. Mechanosensitive channels are transmembrane ion channels regulated by mechanical stimuli \cite{Jin2020}. They are those that transfer external mechanical triggers inside the cells and convert them to biochemical signals, and thus play crucial roles in heat and touch sensations. By addressing the complexity associated with these channels, nano-tweezing capabilities have been utilized to study how small forces affect important domains of mechanosensitive channels and their physiological behaviors. Piezo channels, which are large proteins with propeller-shaped structures \cite{Saotome2017}, are another example of mechanosensitive channels. These review articles \cite{Neuman2008,Wang2022,Yang2022} summarize the applications of optical tweezers in various mechanosensitivity studies.

Fluorescence imaging coupled with dynamic force spectroscopy can directly image and correlate the external force acting on the mechanoreceptors to binding kinetics and biochemical signals \cite{Zhu2019}. For example, the optical tweezers-based technique revealed that the mechanosensor $\alpha \beta TCR$’s response is anisotropic, where its costimulatory activation is triggered  by only a tangential external force, but not by a normal force \cite{Kim2009}. It was also demonstrated that in the absence of any mechanical stimuli, the initial TCR calcium triggering requires a much higher pMHC ligand density \cite{Feng2017}. Calcium mobilization with increased membrane tension was observed and associated with the activation of the TRP-4 channel while pulling membrane tethers from Caenorhabditis elegans DAV neurons using a similar technique \cite{Das2021}. Nano-tweezers have been employed to control mechanosensitive channel behavior by applying membrane deformations in more complex systems, tapping or indenting cell membranes, and studying $Ca^{2+}$ transients \cite{Falleroni2018}.

Mechanical force has also been shown to activate receptor proteins when the cells come in contact, including members of the Notch family. Single-molecule optical tweezers helped to probe the forces in Notch–ligand interactions \cite{MelotyKapella2012,Shergill2012,Stephenson2012}, and finally to conclude that the proteolytic sensitivity increased when a Notch was stretched with a small force, even forces in the order of picoNewton was sufficient to expose the cleavage site \cite{Gordon2015}. Another example of receptor–ligand binding probed using optical traps is the interaction between the von Willebrand factor (VWF) and platelet receptor glycoprotein $Ib\alpha (GPIb\alpha)$, which mediates the attachment of platelets on vessel walls \cite{Ruggeri2007}. The capabilities of optical tweezers in resolving forces were utilized to analyze the single-molecule kinetics of VWF and $GPIb\alpha$ pairs \cite{Kim2010,Kim2015a}. Other surface receptors, such as integrins, cadherins, selectins, and T-cell receptors, have also been probed using optical traps \cite{Chen2017}. In similar mechanosensitivity studies employing optical tweezers, the binding probability and unbinding force between virus-coated colloids and erythrocytes were measured by first bringing them in contact, and then applying a gradually increasing force until they detached \cite{Mammen1996a}. Similarly, the binding strength and activation state of single fibrinogen-integrin pairs have been measured in living cells \cite{Litvinov2002}.

\subsection{Biomedical Devices}

Optical micromanipulation of cells and tissues using laser tweezers and scissors has opened up many possibilities in developmental research and biomedical applications. The potential of laser microsurgery for the isolation of single cells or groups of cells from a cell cluster was identified and employed early on \cite{MeierRuge1976,Leitz2003}. The dissection of cellular organelles and cytoskeletal ﬁlaments \cite{Kumar2006} or even chromosomes \cite{Berns1989,Berns1997}  were also performed. An intricate technical setup named laser capture microdissection has also been developed (LCM) \cite{EmmertBuck1996}. The radiation pressure of a high-intensity laser was used to blow a desired cell in order to isolate it from the tissue \cite{Schuetze1997,Schiitze1997, Schueitze1998}.

Focused laser beams provide a much better tool for cell fusion by transient permeabilization of cell membranes upon short-term laser irradiation and molecular injection (opto-injection). A method for successful cell fusion by bringing two cells in contact using optical traps and cutting the common wall with a few pulses of high-power UV laser microbeams was developed and demonstrated by Steubing \textit{et. al.} \cite{Berns1991}. In the laser-beam-assisted opto-injection technique, a temporary opening, that is, a small hole, is created in the cell membrane by focusing a single ultraviolet pulse on the phospholipid bilayer, and then the foreign species or molecules  are injected or allowed to pass through the hole in a passive manner. Improvements over the conventional glass-capillary technique are in terms of success rates, less harm to the cells, and fine control over the hole size \cite{Greulich1999}. Laser-assisted genetic material transfer has been demonstrated in both animal and human cells \cite{Tsukakoshi1984,Kurata1986,Tao1987}. Although the conventional technique is still applicable to animal cells, it is often not useful for plant cells because of their thicker cell membranes \cite{Weber1988a,Weber1988b,Weber1989}. For example, while working with rapeseed cells, thin capillaries are fragile and cannot penetrate the cell membrane, and on the other hand, thicker capillaries damage the cells. However, photo-injection through laser holes in rapeseed cell membranes and tobacco \cite{Weber1989,Weber1990} have been successfully performed. This method is most commonly used to inject fluorescent biomolecules into the cytosol and nucleus of cells, such as Neuro-2A mouse neuroblastoma cells \cite{Stuhrmann2006}. A fluorescent dye can be taken up through laser-created holes into plasmolyzed cells in less than 5 s.

Similar methods provide an effective way to bring two cells together, thus generating hybrid cells without losing cell functions. This method has been successfully applied to \textit{in-vitro} fertilization. Individual sperms are transported using an optical trap and brought into contact with egg cells \cite{Schopper1999}. Alternatively, a micrometer-sized hole is formed in the zona pellucida and the viscous barrier is opened. Sperm cells then swim unassisted and slip through the hole into its interior, where the fertilization process continues. However, more recent techniques combine laser microbeam-assisted zona drilling with the optical trapping of sperm cells to enhance efficiency \cite{Clementsengewald1996}. \textit{In-vitro} fertilization success rates have been compared between methods without laser support and with laser zona drilling and subzonal insemination assisted by optical tweezers in a model experiment with mouse gametes \cite{Enginsu1995}. After \textit{in-vitro} fertilization, very early embryos are placed in the womb, where thinning of the zona facilitates easy implantation.

Optical tweezer-based techniques have also helped in various studies in biomedical research and development, such as the study of virus-cell adhesion, which is of  considerable practical importance. A microsphere coated with virus particles was brought into contact and its adhesion to the cells was studied using a double optical tweezers system \cite{Mammen1996}. An optical tweezer-based immunosensor was also constructed with these mechanisms in mind \cite{Helmerson1997}. Optical micromanipulation-based techniques have also been applied to neuroscience. Optical tweezers have proven to be excellent tools for the manipulation of whole neurons \cite{TownesAnderson1998,Pine2009} or for indirectly probing synapses and receptors using attached particles . In other applications, tissue sections of early gastric tumors have been investigated \cite{Becker1996,Schuetze1997, Schiitze1997}. The fight between human immune cells and cancer cells has also been observed using optical tweezers \cite{Seeger1991}.

\subsection{Optical Micromotors}
\label{sec:microrotor}

A dielectric particle experiences a scattering force, or in other words, radiation pressure, in the light field of a laser beam. The scattering force pushes the particle along the propagation direction. While it can transport the particle from one point to another, in a more interesting scenario, when there is chirality either in the light field or in the particle geometry, it induces torque and thus rotational motion to the particle. Surface-micromachined structures were fabricated with chiral shapes. These specially designed chiral shapes, often made from transparent polymers using photolithography, exhibit spontaneous rotation, as the scattering force produces a torque owing to the finite chirality of the particles, while the gradient force holds them, providing a hinge. Another way to induce light-driven rotation is to use special beam modes with finite orbital angular momentum to form an optical trap. Using these special beam modes or dynamically changing beam shapes, non-chiral particles can also be rotated under the trap. In an alternative realization, optical tweezers with circularly polarized laser beams have been shown to induce torque on the birefringent particles. This light-induced rotation is of great importance because of its potential applications in optically driven micro-machines, motors, actuators, and biological specimens.

The first demonstration of an optically driven micromotor was realized by optically trapping 10–25 $\mu m$ diameter silicon dioxide ($SiO_2$) structures made by ion-beam etching \cite{Higurashi1994}. Other examples include surface-micromachined $SiO_2$ rotors \cite{Friese2001} and optically fabricated microstructures with various shapes (helices, sprinklers, and propellers) under optical traps \cite{Galajda2001}. Synthesized inorganic nanorods have been shown to rotate when trapped in optical tweezers, where their random surface roughness provides finite chirality \cite{Khan2006,Khan2007}. Even deformed human erythrocytes exhibit rotation under an optical trap because of their asymmetric chiral shapes upon crenation in a hypertonic buffer \cite{Khan2005,Khan2009}. Asymmetric non-chiral particles, such as rods and ellipsoids, often exhibit rotational motion when trapped in certain configurations where the asymmetric scattering force generates a torque \cite{Khan2006,Khan2007,Zong2015, Petkov2017}. Symmetric particles with asymmetric scattering profiles, such as metal half-coated spherical Janus colloids, also experience torque owing to an imbalance in the optical forces acting on the two hemispheres \cite{Zong2015}.

Optical traps with circularly polarized light, which carries spin angular momentum, are commonly used to induce spinning of trapped birefringent particles. This method has been employed for the rotation of elongated particles and a cluster of particles, in addition to birefringent particles \cite{Bradac2018}. However, the maximal torque experienced by the trapped object is much higher in the case of a birefringent particle than in the case of an elongated particle \cite{FernandezNieves2005}. Rotation rates as high as 350 Hz and 10 MHz were achieved using a circularly polarized laser beam for an optically trapped micron-sized birefringent particle in water \cite{Harada1996} and vacuum \cite{Stilgoe2008}, respectively. Selective 3D trapping of chiral micro- and nanoparticles with circularly polarized light, where the focused beam can induce non-restoring or restoring forces on the chiral particle, has also been demonstrated \cite{Tkachenko2014,Vovk2017}. Deformed and crenated human erythrocytes also exhibit rotation under a circularly polarized optical trap, the rate of rotation (at constant laser power) being a measure of their deformation \cite{Dharmadhikari2004,Ghosh2006,Bambardekar2008}.

A structured laser beam carrying angular momentum also causes rotation of the trapped micro- and nano-objects because of the transfer of angular momentum. The concept of a light beam possessing orbital angular momentum, which is an azimuthally varying phase, was first introduced in 1992 \cite{Allen1992}. However, the use of such a laser beam in an optical trap, which induces rotation to a trapped absorptive dielectric particle, was first demonstrated by the group of Rubinsztein–Dunlop \cite{He1995,He1995a}. LG modes of circularly polarized light, carrying both orbital and spin angular momentum, have also been used to induce rotation on absorptive and birefringent particles about their own axis as well as the beam axis \cite{Friese1996,ONeil2002,GarcesChavez2003}. In this case, the rotation rate is proportional to the sum of the orbital and spin angular momenta \cite{Parkin2006}. This method has been used to study active microrheology, where the spinning of a wax disk applies torque to the surrounding viscoelastic medium and measures its response \cite{Wilking2008}.

A laser beam with an externally modulated time-dependent intensity profile can form a dynamic trap, causing the trapped particles to move along the desired direction or rotate about a desired axis and rate. Following this concept, experimental studies have been carried out with the optical realization of a thermal ratchet where the laser beam intensity is modulated asymmetrically along a circle, causing circular motion of micron-sized particles trapped in the rim \cite{Faucheux1995}. Similar experimental realizations with externally modulated optical traps include Feynman’s ratchet with colloidal particles (1D) \cite{Bang2018}, reconfigurable two-dimensional (2D) rocking ratchet systems \cite{Arzola2017}, and feedback-controlled flashing ratchets \cite{Lopez2008}. Transport of trapped microparticles has been achieved in a tilted line (one-dimensional) trap; thus, by applying an asymmetric gradient force \cite{Liesfeld2003,Khan2006a,Liu2016,Shen2022}. Particle transport can also be realized using bistable, circular, sinusoidally oscillating optical traps, and pattern holographic tweezers \cite{Faucheux1995,Gahagan1996,McCann1999,Liesfeld2003,Lutz2004,Mu2009, Liu2016, Shen2022}.

While most noise-driven micro-machines, such as thermal ratchets and Brownian motors, rely on time-dependent forces, time-independent non-conservative force fields can also create stochastic toroidal motion of trapped particles, called Brownian vortexes. When dielectric particles diffuse inside radially symmetric Gaussian profiled laser beams, they also experience scattering forces that are nonconservative in nature. The trapped particles thus experience circulatory bias and exhibit toroidal circulation inside the trap owing to the non-conservative solenoid component of the optical forces \cite{Roichman2008,Sun2009}. Brownian vortices at the liquid-air interface have also been demonstrated by applying an external non-conservative drag force to a colloidal particle bound by optical tweezers \cite{Khan2011}.

\subsection{Microfluidics Devices}

In many studies and applications, microfluidic channels and devices have been proven to be immensely beneficial because they enable working with small sample volumes (micro-liters to nanoliters) and to fulfill special requirements, such as setting up customized flow patterns or maintaining desired ionic environments with the help of a buffer that needs to be in proximity yet separated by a semipermeable barrier. Another advantage of working with flows in a microfluidic channel is that in narrow channels, the flow is always in the low Reynolds number regime where viscosity dominates, and inertial effects may largely be ignored. In the absence of turbulence, mixing and other related phenomena are mostly governed by diffusion, which also slows down inside the narrow channels. This brings in interesting new applications, besides making it easier to observe flows under a microscope. Custom-designed micro-channels or microfluidic devices can easily be fabricated as per customized design with optically transparent material (such as PDMS film) using electron beam or laser-based lithography techniques \cite{Niculescu2021}. 

Optical forces can be conveniently used to drive and actuate micro-components, such as micro-cogs \cite{Galajda2001} and micro-rotors (please see section \ref{sec:microrotor}), which are rotated by exploiting the spin or angular momentum of the light beam or the chiral geometry of the rotor, to set up small pumps. Specific realizations include an optically driven microfluidic pump driven by two rotating birefringent vaterite particles in linearly polarized optical tweezers \cite{Leach2006}. An optical trap was integrated with a microfluidic chamber by placing the trapping laser adjacent to the flow channel to make a fully portable and integrated lab-on-chip device for various applications \cite{CranMcGreehin2006}. 

Sorting cells, colloidal particles, or other microscopic objects is another important application. The easy integration of optical tweezers with microfluidic devices for single-cell manipulation has been demonstrated \cite{Enger2004,Dholakia2006,Monat2007,Zhang2008}. Single-cell synthesis in microfluidic channels using optical manipulation results in homogeneous daughter cells, which is a significant enabler in cell therapy. Optical trap-integrated microfluidic devices are also used to study the response of a cell to its neighboring cells, other biomaterials or surfaces, and external chemical, mechanical, and electrical stimuli. High-throughput, automated, passive cell sorting in microfluidic channels has been proven to be more convenient than the conventionally used industry standard - fluorescence-activated cell sorter based on a flow cytometer. The new separation methods based on optical tweezers do not require fluorescent tagging because they distinguish cells based on their individual physical properties. Here, an optical force was employed to trap the cells of interest and for further recognition, sorting, transportation, and isolation from the batch \cite{Terray2002,Applegate2004,Paterson2005,Wang2005,Wang2011,Paie2018}. Thus, this technique can achieve a very high throughput, which can be further enhanced by forming an extended 2D or 3D optical potential landscape. Further improvements are realized by using a scanning mirror \cite{Wang2020}, sharing the beam between multiple traps by acoustic-optical deflectors (AODs) \cite{Endres2016}, and creating extended trap configurations with a spatial light modulator (SLM) \cite{Chapin2006}. The optical scattering force is also useful for sorting cells by pushing them along the laser propagation direction. We have already discussed the process of characterizing erythrocyte deformability by their spinning speed under an optical trap \cite{Dharmadhikari2004,Khan2005,Ghosh2006,Bambardekar2008,Khan2009}, which can be performed more conveniently in a microfluidic channel.  

Optical trap-assisted particle sorting and separation in a microfluidic chamber, that is, optical fractionation, has also been performed by tuning the competition between the attractive potential provided by the trap, which depends on the particle size, refractive index, and laser power and Stokes drag, which can be externally regulated by the flow speed \cite{Zemanek2002,MacDonald2003}.  Such sorting or separation offers new potential for the selection and enrichment of cell phenotypes. More recently, holographic techniques and evanescent waves have been employed to create an array of optical traps \cite{Mellor2006} and in another case, near-field optics have been used to enhance the plasmonic field \cite{GarcesChavez2006} to simultaneously manipulate a large number of particles in microfluidic devices. Holographic-based tweezers, which provide the capability to generate dynamically programmable trap patterns, make it an easy tool for automated particle capture and sorting \cite{Chapin2006}.

\section{Conclusion}

This chapter covers the theory of optical micromanipulations, design considerations of conventional single-trap and dual-trap systems, as well as recent improvements and adaptations. Furthermore, the diverse capabilities of this technique and some of its widely used applications have been discussed. Nevertheless, it is important to acknowledge that this chapter does not aim to provide an exhaustive exploration of this extensive field of research and application. There have been boundless adaptations and customizations of optical micromanipulations, which, as a whole, are beyond the scope of this chapter. Instead, this is an introductory overview of the advantages and possibilities of optical tweezers for investigating functional soft materials and designing application-oriented devices.

Optical micromanipulation has emerged as one of the most convenient and versatile tools for studying soft matter systems because of its ability to hold and manipulate micron and sub-micron objects in a noninvasive and nonmechanical manner. These capabilities make them particularly useful for working with biological samples such as living cells. Laser scissors and tweezers are used in various challenging biomedical applications, including laser-based surgeries. By exploiting the ability to measure calibrated forces at the pico-and femtonewton levels, optical micromanipulations are used for in vitro studies of mechanical responses and mechanosensitivity. The structural and viscoelastic properties of complex fluid systems are also commonly investigated using optical traps. In addition, complex and customized optical potential landscapes, such as holographic optical tweezers and time-shared optical trap patterns, are gaining popularity for investigating the statistical physics of soft, active, and biological systems.

The most crucial aspect of optical micromanipulation is choosing an appropriate laser beam depending on the study or application, as the fundamental physical phenomenon is light–matter interaction. The wavelength of the laser not only determines the trapping efficiency but also the extent of scattering, absorption, and photonic damage of the sample. Trapping energy, and hence the efficiency, increases with the photon energy; thus, with the inverse of wavelength. However, the use of shorter wavelengths is not always recommended to enhance trapping efficiency. For particles that are predominantly optically transparent, it is more convenient to operate them in the Mie regime, that is, with longer wavelengths. Moreover, the scattering force pushes the trapped particle along the beam propagation direction, thereby destabilizing the trap. Hence, it is always intended to minimize the scattering and maximize the transmission of the trapping laser to achieve a stronger trap. The absorption of the trapping laser by the trapped microparticle or the surrounding medium results in an increase in the local temperature. Raising the local temperature amplifies the spontaneous thermal fluctuations and induces convective fluid flow around the trapped particle, affecting the stability of the trap. In addition, increased temperatures often ruin biological samples beyond recovery. Another important consideration regarding the trapping wavelength is the photonic damage of the sample, such as the wavelength-dependent photo-induced breakdown of chemical bonds and the consequent degradation of the sample or medium being studied. Occasionally, pulsed laser beams are also utilized for optical micromanipulations to reduce heating, in contrast to continuous-wave laser beams, which are traditionally employed in optical tweezers. However, the required peak power of the laser pulses to achieve a trapping efficiency similar to that of a continuous wave laser must be within the damage threshold of the sample or the optical setup.

Although optical micromanipulation-based diagnostics can detect diseases at an early stage of infection by analyzing the alteration of the physical or optical properties of the infected cells using a small sample volume, biochemistry-based tests continue to be the established benchmarks in the industry. Therefore, to design optical micromanipulation-based devices that are both convenient and economically viable for biomedical applications, the primary considerations are cost effectiveness and portability. The key to building more compact and popular biomedical devices lies in the design and use of high-quality small-diode lasers and adaptive yet cost-effective optical elements constructed from responsive soft materials.

\bibliography{OT_SoftMatter}

\end{document}